\documentclass[11pt,a4paper]{article}
\pdfoutput=1

\usepackage[T1]{fontenc}
\usepackage{amsmath}
\numberwithin{equation}{section}
\usepackage{amsthm}
\usepackage{amsfonts}
\usepackage{amssymb}
\usepackage{graphicx,subfigure,color}
\usepackage{hyperref}
\usepackage[font=footnotesize,labelfont=bf]{caption}
\usepackage{amscd}
\usepackage{enumerate}
  \usepackage{cite}

\def\ba{\begin{eqnarray}}
\def\ea{\end{eqnarray}}
\def\w{\wedge}

\begin{document}
\def\thefootnote{\fnsymbol{footnote}}

\begin{center}
\Large{\textbf{Anisotropic inflation with a non-minimally coupled electromagnetic field to gravity}} \\[0.3cm]

\large{Muzaffer Adak$^{\rm a}$, \"{O}zg\"{u}r Akarsu$^{\rm b}$, Tekin Dereli$^{\rm c}$, \"{O}zcan Sert$^{\rm d}$}
\\[0.3cm]

\small{
\textit{$^{\rm a}$ Department of Physics, Pamukkale University, 20017  K{\i}n{\i}kl{\i}, Denizli, Turkey}}

\small{
\textit{$^{\rm b}$ Department of Physics, {\.I}stanbul Technical University, 34469 Maslak,  {\.I}stanbul, Turkey}}

\small{
\textit{$^{\rm c}$ Department of Physics, Ko\c{c} University, 34450 Sar{\i}yer, {\.I}stanbul, Turkey}}

\small{
\textit{$^{\rm d}$ Department of Mathematics, Pamukkale University, 20017  K{\i}n{\i}kl{\i}, Denizli, Turkey}}

\end{center}

\vspace{.5cm}

\hrule \vspace{0.2cm}
\noindent \small{\textbf{Abstract}\\
We consider the non-minimal model of gravity in $Y(R) F^2$-form. We investigate a particular case of the model, for which the higher order derivatives are eliminated but the scalar curvature $R$ is kept to be dynamical via the constraint $Y_RF_{mn}F^{mn} =-\frac{2}{\kappa^2}$. The effective fluid obtained can be represented by interacting electromagnetic field and vacuum depending on $Y(R)$, namely, the energy density of the vacuum tracks $R$ while energy density of the conventional electromagnetic field is dynamically scaled with the factor $\frac{Y(R)}{2}$. We give exact solutions for anisotropic inflation by assuming the volume scale factor of the Universe exhibits a power-law expansion. The directional scale factors do not necessarily exhibit power-law expansion, which would give rise to a constant expansion anisotropy, but expand non-trivially and give rise to a non-monotonically evolving expansion anisotropy that eventually converges to a non-zero constant. Relying on this fact, we discuss the anisotropic e-fold during the inflation by considering observed scale invariance in CMB and demanding the Universe to undergo the same amount of e-folds in all directions. We calculate the residual expansion anisotropy at the end of inflation, though as a result of non-monotonic behaviour of expansion anisotropy all the axes of the Universe undergo the same of amount of e-folds by the end of inflation. We also discuss the generation of the modified electromagnetic field during the first few e-folds of the inflation and its persistence against to the vacuum till end of inflation. } 
\\
\noindent
\hrule
\def\thefootnote{\arabic{footnote}}
\setcounter{footnote}{0}
\let\thefootnote\relax\footnote{\textbf{E-Mail:} madak@pau.edu.tr, akarsuo@itu.edu.tr, tdereli@ku.edu.tr, osert@pau.edu.tr}
\def\thefootnote{\arabic{footnote}}
\setcounter{footnote}{0}

\section{Introduction}

Inflationary cosmology \cite{Starobinsky:1980te,Guth:1980zm,Linde:1981mu,Albrecht:1982wi}, the epoch of quasi-de Sitter expansion in the very early Universe, relieves the standard Big Bang cosmology from the requirement of extremely finely tuned initial conditions. The simplest examples based on a single scalar field generically predict statistically isotropic, Gaussian and almost scale invariant Cosmic Microwave Background (CMB) temperature fluctuations  (see \cite{Linde:2014nna} for a recent review). Inflation together with $\Lambda$ cold dark matter leads to the minimal inflationary $\Lambda$CDM model. Analyses of the data from CMB \cite{Hinshaw:2012aka,Ade:2015xua}, type Ia supernovae (SNIa) \cite{Betoule:2014frx} and baryonic acoustic oscillations (BAO) \cite{Aubourg:2014yra} observations allow us not only to fit all free parameters of this model with high accuracy, but also to test its underlying assumptions. The high precision data from WMAP \cite{Hinshaw:2012aka} and Planck \cite{Ade:2015xua} satellites agree with this model but reveal several unexpected features at large angular scales \cite{Schwarz:2015cma} that might be addressed in the context of anisotropic inflation.

Nature of primordial fluctuations that lead to aforementioned features of the CMB can be understood by symmetries during inflation \cite{Baumann:2009ds}. The \textit{shift symmetry} in scalar field space, due to the sufficiently flat potential required for slow-roll inflation, implies suppression of non-linearity and hence statistically Gaussian fluctuations. The non-Gaussianity predicted by single field slow-roll inflation is negligibly small so that it cannot be observed in CMB observations. The \textit{temporal de Sitter symmetry} ($\dot{H}=0$) yields a scale invariant power spectrum because of the scale invariance of spatial coordinates. Sufficiently long but finite stage of inflation requires violation of this symmetry which implies a deviation from scale invariant power spectrum characterised by a scalar potential as $\dot{H}=\frac{\partial_{\phi} V(\phi)}{2\sqrt{3V(\phi)}}$. The \textit{spatial de Sitter symmetry} forbidding direction dependence due to the rotational symmetry ($\sigma^2=0$) leads to statistically isotropic fluctuations. Wald's cosmic no-hair theorem, which follows the cosmic no-hair conjecture arguing that the late-time behaviour of any accelerating Universe is an isotropic Universe given first in \cite{Gibbons:1977mu,Hawking:1981fz}, states that Bianchi-type models (except Bianchi IX) in the presence of a positive cosmological constant and a source satisfying strong and dominant energy conditions will approach de Sitter space exponentially fast, within a few Hubble times \cite{Wald:1983ky}. Afterwards, it has been long thought that local\footnote{The Wald's theorem \cite{Wald:1983ky} and results of the papers \cite{Moss:1986ud,Kitada:1991ih} consider the spatially homogeneous spacetimes only. A generic late-time attractor in general theory of relativity with a cosmological constant first obtained in \cite{Starobinsky:1982mr} has inhomogeneous time-independent tensor hairs outside the future de Sitter event horizon which may have an arbitrarily large amplitude. Its generalization to the case of power-law inflation presented in \cite{Muller:1989rp} has time-independent scalar hairs in addition. Hence, though one may say that the cosmic no-hair conjecture is valid locally, i.e., for one observer only, it is not correct globally. In fact, these hairs even become observable after the end of inflation, and can be directly seen today in fluctuations of CMB temperature and polarization.} isotropy is the most robust prediction of scalar field inflaton models \cite{Moss:1986ud,Kitada:1991ih}. There is currently no observational evidence for non-Gaussianity \cite{Ade:2015ava} but for a slight deviation from scale invariance ($n_s=1$), which implies $\dot{H}<0$ during inflation and hence $n_s<1$, with more than $5\sigma$ \cite{Ade:2015lrj}. These are in line with the single scalar field slow-roll inflation. However, the lack of angular correlation at large angles, alignments of the lowest multipole moments, and hemispherical power asymmetry all together constitute strong evidence that broken isotropy is a real feature of CMB (see \cite{Schwarz:2015cma} and references therein). This might be addressed by violating the rotational symmetry during inflation, i.e., anisotropic inflation that challenges the cosmic no-hair conjecture. In fact, there have been challenges to the cosmic no-hair conjecture earlier \cite{Ford:1989me,Kaloper:1991rw,Kawai:1998bn,Barrow:2005qv,Ackerman:2007nb,Golovnev:2008cf,Kanno:2008gn}, albeit they were suffering from several problems such as instability \cite{Himmetoglu:2008zp,EspositoFarese:2009aj,Golovnev:2009rm}. It is recently showed in Ref. \cite{Watanabe:2009ct} that if a canonical scalar field $\phi$ with a potential $V(\phi)$ couples to a vector kinetic term $F_{\mu\nu}F^{\mu\nu}$, the vector field can be persistent, and hence leads to anisotropic hair during inflation for a suitable choice of the coupling $f^2({\phi})$. It is then showed in Ref.\cite{Kanno:2010nr,Hervik:2011xm} that there exists an anisotropic power-law inflationary attractor yielding constant expansion anisotropy $\sigma^2/H^2$, in the presence of exponential scalar potential and such a coupling of the exponential form, which often appears in string theory and supergravity \cite{Ratra:1991bn,Martin:2007ue}. These studies are followed by many cosmological predictions \cite{Ackerman:2007nb,Pereira:2007yy,Pitrou:2008gk,Watanabe:2010fh,Soda:2012zm,Kim:2013gka,Ito:2016aai} and further extension to various models \cite{Do:2011zza,Do:2011zz,Bhowmick:2011em,Thorsrud:2012mu,Maleknejad:2012fw,Ohashi:2013mka,Ohashi:2013pca,Akarsu:2013dva,Koivisto:2015vda,Ito:2015sxj,Sundell:2015gra,Do:2016ofi,Fukushima:2016wyz,Lahiri:2016jqv,Heisenberg:2016wtr}.

The existence of cosmological scalar fields has not yet been established observationally and they are mostly introduced in an ad hoc way (see \cite{Martin:2013tda} for a comprehensive list). The gauge fields, which yield anisotropic stress, on the other hand are far more common at all energy scales, in particular at energies relevant to inflation. Any scalar field model, canonical or not, can in principle be represented by an interacting vacuum, viz. dynamical cosmological ``constant'' that would mimic scalar potential, plus an isotropic fluid that would mimic scalar kinetic term, and vice versa (see for instance \cite{DeSantiago:2012xh}). Replacing the isotropic source here with a vector field would thus not only lead to an alternative to non-minimally coupled scalar and vector fields models but also rolling scalar field as the inflaton. Namely, in this case the vector field that would be persistent by gaining energy-momentum from the vacuum can lead to anisotropic hair, while the dynamical vacuum, due to the  persistent vector field, leads to a temporal deviation from de Sitter expansion. It is indeed showed in \cite{Maleknejad:2012as} that in the presence of a generic anisotropic fluid described by an energy-momentum tensor of the form of a dynamical cosmological ``constant'', i.e., vacuum, plus a piece which does not respect strong or dominant energy conditions, expansion anisotropy may grow in spite of the accelerated expansion. The point in such a setting is that the broken scale invariance $\dot{H}<0$ implies broken statistical isotropy $\sigma^2\neq 0$, such that if the vacuum is not interacting then it would yield constant energy density so that Wald's cosmic no-hair theorem would be valid and hence neither the scale invariance nor the isotropy would be broken. This is in line with that the broken isotropy seems to happen only at the largest angular scales, it also amounts to a violation of scale invariance \cite{Schwarz:2015cma}. Such a source may appear as an effective energy-momentum tensor from the non-minimal coupling of electromagnetic field to gravity, and may lead to generalisations of the Einstein-Maxwell model as was first done in \cite{Prasanna71}. However, variations of an arbitrary Lagrangian with respect to non-minimally coupled gravitational and electromagnetic fields yield complicated field equations which in general may involve higher derivatives \cite{Balakin:2005fu}, and hence one needs to introduce simplifying assumptions for further physical investigations. For instance,  in Ref. \cite{Horndeski:1976gi}, a $\frac{R}{M^2}F^2$-type model \cite{Horndeski:1976gi,Turner:1987bw,MuellerHoissen:1988bp,Balakin:2010ar,Dereli:2011hu,Baykal:2015paa} with the field equations involving only the second order derivatives of the variables is constructed and it is showed that the Einstein-Maxwell field equations with a cosmological constant appears as a special case of the model. As another example, $\frac{R^\eta}{M^{2\eta}} F^2$-type models \cite{Mazzitelli:1995mp,Bamba:2008ja,Campanelli:2008qp,Lambiase:2008zz,Kunze:2009bs,Dereli:2011gh}, where $\eta$ is a positive integer, are considered for the generation of seed magnetic fields during the inflation to explain the origin of large-scale magnetic fields present in all galaxies \cite{Turner:1987bw,Mazzitelli:1995mp,Campanelli:2008qp,Lambiase:2008zz}. Although the non-minimal coupling of the electromagnetic field is classically forbidden by the equivalence principle, they arise in the effective Lagrangian in various theories. For example, in photon propagation in quantum electrodynamics in a curved background \cite{Drummond:1979pp} and in the dimensional reduction of Gauss-Bonnet and the general curvature-squared terms in higher dimensions \cite{Buchdahl:1979wi,MuellerHoissen:1987ch,Dereli:1990he}. The non-minimal model of gravity in $Y(R)F^2$-form we studied in this paper is characterised by the constraint $F_{mn}F^{mn} \,  \frac{{\rm d} Y}{{\rm d} R} =-\frac{2}{\kappa^2}$, which eliminates the higher order derivatives in the field equations and renders $R$ dynamical. This constraint intertwines the an/isotropisation of the Universe, the deviation from the temporal de Sitter symmetry (which is inevitable in our model) and the generation/persistence of the electromagnetic field with each other in a particular way. We study the cosmology of the model in the presence of magnetic field only, i.e., under the assumption $F_{mn}F^{mn}>0$. Our model admits non-minimal coupling in $\frac{R^\eta}{M^{2\eta}} F^2$-form, but in a particular way as $\frac{\eta}{M^{2\eta}} R^{\eta-1}  F_{mn}F^{mn}=-\frac{2}{\kappa^2}$ with $\eta<0$ in contrast to the similar models we mentioned above. In this case, the spectral index of curvature perturbations (a measure of the deviation from the temporal de Sitter symmetry), the expansion anisotropy and the ratio of the energy density of the modified electromagnetic field to the total energy density of the Universe appear as constant functions of $\eta$.

The paper is organised as follows. In Section \ref{model}, motivated by the above discussions, we consider the non-minimal model of gravity in $Y(R) F^2$-form \cite{Bamba:2008xa,Dereli:2011mk}. We investigate a particular case of the model, for which the higher order derivatives are eliminated but $R$ is kept to be dynamical rather than choosing $R=0$ that would lead to nothing new but the Einstein-Maxwell model. We showed that our model is equivalent to general theory of relativity (GR) in the presence of a single fluid source that can be represented by interacting electromagnetic field and vacuum depending on $Y(R)$, namely, the energy density of the vacuum tracks the scalar curvature while energy density of the conventional electromagnetic field is dynamically scaled with the factor $\frac{Y(R)}{2}$. We discussed also that modified electromagnetic field can be generated and become persistent during inflation. In Section \ref{sec:cosmo}, we construct an anisotropic cosmological model, and then discuss the dynamics of the modified electromagnetic field and vacuum as well as the resultant anisotropic effective fluid, and the expansion kinematics of the Universe. In Section \ref{sec:plve}, we give exact solutions for anisotropic inflation by assuming the volume scale factor of the Universe exhibits a power-law expansion rather than each of the scale factors in contrast to the models given, for instance, in \cite{Kanno:2010nr,Do:2011zza,Do:2011zz,Bhowmick:2011em,Ohashi:2013pca,Do:2016ofi}. We find that directional scale factors do not necessarily exhibit power-law expansion, which would give rise to a constant expansion anisotropy, but expand non-trivially and give rise to a non-monotonically evolving expansion anisotropy that eventually converges to a non-zero constant. Relying on this fact, we direct our attention to the anisotropic e-fold during the inflation by considering observed scale invariance in CMB and demanding the Universe to undergo the same amount of e-folds in all directions as in the isotropic inflationary models based on RW spacetimes. Considering the observed value of the spectral index that measures the deviation from scale invariance, we calculate the residual expansion anisotropy at the end of inflation, though as a result of non-monotonic behaviour of expansion anisotropy all the axes of the Universe undergo the same of amount of e-folds by the end of inflation. We further discuss the generation and persistence of the modified electromagnetic field during the inflation. In Section \ref{section:CR}, we point out our closing remarks and discuss the future perspectives.

\section{A non-minimally coupled electromagnetic field to gravity}
\label{model}

The field equations to be considered in this paper are obtained from the non-minimal model of gravity in $Y(R) F^2$-form by a variational principle from the action
\begin{equation}
I[{\omega^a}_b,e^a,F] = \int_\mathcal{M}{L},
\end{equation}
where we integrate the Lagrangian over a four-dimensional manifold  $\mathcal{M}$ that has the metric $g = \eta_{ab} e^a \otimes e^b$ with $\eta_{ab}=\mbox{diag}(-+++)$. Here $e^a$ are the orthonormal basis 1-forms and $\omega_{ab}=-\omega_{ba}$ are the corresponding Levi-Civita connection 1-forms. $F={\rm d}A$ is the electromagnetic field 2-form. The orientation of the manifold is fixed by the Hodge map $*1 = e^0 \w e^1 \w e^2 \w e^3$. Spacetime curvature tensor 2-forms  ${R^a}_b$ and torsion tensor 2-forms $T^a$ are determined from the structure equations:
\begin{align}
\label{Rba}
R^{a}_{\;\;b} = {\rm d}\omega^{a}_{\;\;b} + \omega^{a}_{\;\;c} \w \omega^{c}_{\;\;b} \, ,
\\
T^a = {\rm d}e^a + \omega^{a}_{\;\;b} \w e^b \ .   
\end{align}
We consider the following Lagrangian density 4-form \cite{Bamba:2008xa,Dereli:2011mk}:
\begin{equation}
\label{lag1}
L =  \frac{1}{2\kappa^2} R*1 -\frac{1}{2}Y(R) F\w *F + \lambda_a  \w  T^a  + \mu \wedge d F \ ,
\end{equation}
where $\kappa^2 = 8\pi G$ is Newton's universal gravitational constant $(c=1)$, $\mu$ is a Lagrange multiplier 2-form that leads to homogeneous electromagnetic field, i.e., ${\rm d}F=0$ or $F={\rm d}A$, $\lambda_a$ are Lagrange multiplier 2-forms that lead to the torsion-less case $T^a=0$. The curvature scalar is denoted by $R$ and $Y(R)$ at this point is any function of $R$. The electromagnetic field 2-form can be expanded as $F = \frac{1}{2} F_{ab} e^a  \w  e^b$. Throughout this paper we use the interior product operators $\iota_a\equiv \iota_{X_a}$ which satisfy $\iota_a e^b= e^b(X_a)  = \delta^b_a$ and the shorthand notation $ e^a \wedge e^b \wedge \cdots = e^{ab\cdots}$. We define the Ricci 1-forms $R_b= \iota_a {R^a}_b$, the curvature scalar $\iota_{ba} R^{ab}= R $ and $\iota_aF =F_a$,  $\iota_{ba} F =F_{ab} $. 

The infinitesimal variations of the total Lagrangian density $L$ w.r.t. $\{e^a\}$, ${\{\omega^a}_b\}$ and $A$, up to a closed form, are given by
\begin{align}\label{generaleinsteinfe1}
   \delta{L}  = \frac{1}{2 \kappa^2} \delta{e}^a \w R^{bc} \w *e_{abc}
     + \delta{e}^a \w \frac{1}{2} Y(R) (\iota_a F \w *F - F \w \iota_a *F)  + \delta{e}^a \w {\rm D} \lambda_a  \nonumber \\
    + \delta{e}^a \w  Y_R (\iota_a R^b)\iota_b(F \w *F)
      + \frac{1}{2} \delta\omega_{ab} \w ( e^b \w \lambda^a - e^a \w \lambda^b) \nonumber \\
 \delta{\omega}_{ab} \w  {\Sigma}^{ab} -\delta{ F} \w Y(R)  *F   + \delta{\lambda}_a \w T^a
     + \delta{F} \w d\mu + \delta{\mu} \wedge dF   \,,
\end{align}
where  $Y_R = \frac{{\rm d}Y}{{\rm d}R}$, and the angular momentum  tensor is given by
\begin{equation}\label{sigmaab1}
 {\Sigma}^{ab} =  \frac{ 1}{2} {\rm D}   [\iota^{ab}( Y_R F  \w *F )]\ .
   \end{equation}
Solving $\lambda_a$ from the connection equation and then using it in the co-frame field equation of the model we reach the following modified Einstein field equations
\begin{eqnarray}\label{einstein}
\frac{1}{2 \kappa^2}  R^{bc} \w *e_{abc}+  \frac{1}{2} Y(\iota_a F \w *F - F \w \iota_a *F)
+ \frac{Y_R}{2}F_{mn}F^{mn} * R_a 
+ \frac{1}{2}  {\rm D} [ \iota^b {\rm d}(Y_R F_{mn} F^{mn} )]\wedge *e_{ab}= 0  \ ,
\end{eqnarray}
and the modified Maxwell equations
\begin{equation}\label{maxwell1}
{\rm d}F=0 \quad \textnormal{and}\quad {\rm d}(Y * F) = 0  \ .
\end{equation}

Considering Einstein's general theory of relativity, namely $G_a=\kappa^2\tau_a$, where $G_a=-\frac{1}{2} R^{bc} \wedge * e_{abc}=*R_a - \frac{1}{2} R  *e_a$ is the Einstein tensor and  $\tau_a$ is the energy-momentum tensor (EMT), \eqref{einstein} can be written as follows:
 \begin{eqnarray}\label{einstein3}
*R_a - \frac{1}{2} R  *e_a = \kappa^2\, \left\{ \frac{1}{2} Y(\iota_a F \w *F - F \w \iota_a *F)
+ \frac{Y_R}{2}F_{mn}F^{mn} * R_a 
+ \frac{1}{2}  {\rm D} [ \iota^b {\rm d}(Y_R F_{mn} F^{mn} )]\wedge *e_{ab}\right\} \ .
 \end{eqnarray} 
According to this, the model under consideration may be given in an equivalent way if we introduce the following effective EMT in GR:
\begin{equation}
\label{effemt}
\tau_a= \frac{1}{2} Y(\iota_a F \w *F - F \w \iota_a *F)
+ \frac{Y_R}{2}F_{mn}F^{mn} * R_a 
+ \frac{1}{2}  {\rm D} [ \iota^b {\rm d}(Y_R F_{mn} F^{mn} )]\wedge *e_{ab}
\end{equation}
that is subject to the modified Maxwell equations \eqref{maxwell1}, and satisfies the conservation law of EMT, ${\rm D} \tau_a=0$, automatically, since the twice-contracted second Bianchi identity for the Einstein tensor vanishes: ${\rm D} G_a=0$. We note that $\tau_a$ given by \eqref{effemt} involves higher order derivatives that make the model too complicated for a general analytical investigation. Therefore we will proceed with a particular case of the model for which the higher order derivatives are eliminated by the introduction of the following constraint:
 \begin{equation}
 \label{condition0}
Y_RF_{mn}F^{mn} =C={\rm constant}.
 \end{equation}
One may check from \eqref{maxwell1} and \eqref{einstein3} that the choice $Y=1$, which leads to $C=0$, reduces the model to the well known minimally coupled Einstein-Maxwell model, i.e., the effective EMT reduces to the conventional electromagnetic field EMT $\tau_a^{({\rm em})}= \frac{1}{2} (\iota_a F \w *F - F \w \iota_a *F)$, which yields scalar curvature equal to zero $R=0$ since $\tau_a^{({\rm em})}$ is traceless. Generalisation to  $Y={\rm constant}$ changes nothing but leads to a rescaled electromagnetic field EMT by a constant factor $Y$ for the effective EMT as $\tau_a=Y\tau_a^{({\rm em})}$. On the other hand, we are interested in the cases where $R$ is not necessarily null but fulfills the condition \eqref{condition0}. Accordingly, using \eqref{condition0} and taking the trace of the Einstein field equations \eqref{einstein3}, we find
\begin{equation}
\left(1+\kappa^2\frac{C}{2}\right)R=0,
\end{equation}
which is satisfied by the following two solutions
\begin{eqnarray}
\label{Cs}
C=- \frac{2}{\kappa^2 }\quad \textnormal{or}\quad R=0.
 \end{eqnarray}
Thus, considering the former solution, we reach a model described by the following modified Einstein field equations
 \begin{eqnarray}\label{einstein2}
*R_a - \frac{1}{2} R  *e_a= \kappa^2\frac{Y}{2} (\iota_a F \w *F - F \w \iota_a *F) - * R_a   \,
 \end{eqnarray} 
subject to the constraint
 \begin{eqnarray}
 \label{conditionX}
 F_{mn}F^{mn} \,  \frac{{\rm d} Y}{{\rm d} R} =-\frac{2}{\kappa^2}
 \end{eqnarray} 
and the modified Maxwell equations \eqref{maxwell1}.

We imposed the constraint \eqref{conditionX} on the general non-minimal model of gravity in $Y(R)F^2$-form that leads to the field equations \eqref{einstein2}, by eliminating the higher order derivatives in the general field equations \eqref{einstein3}, and to the scalar curvature $R$ which is not necessarily null. The model we further investigate in what follows is characterised by this constraint \eqref{conditionX} and hence it would be useful to discuss here some features and consequences of it. We should first note that, in our conventions, the term $F_{mn}F^{mn}$ is positive, null and negative for magnetic field, electromagnetic radiation and electrical field, respectively. And in this study we consider only magnetic field and hence in what follows we carry out our discussions by taking $F_{mn}F^{mn}>0$, which in turn implies from the condition \eqref{conditionX} that $\frac{{\rm d}Y}{{\rm d}R}<0$. Next, we are particularly interested in obtaining inflationary cosmological solutions, which would be expected to satisfy at least the following two conditions $R>0$ and $\dot{R}\sim 0$, i.e., not to deviate much from the de Sitter inflation (which yields $R=4\Lambda$ and $\dot{R}=0$, where $\Lambda$ is a positive cosmological constant). This is in line with that we leave the case $R=0$ out of our discussion and continue with the case that renders $R$ variable via the constraint \eqref{conditionX} and hopefully fulfils the requirements of inflationary cosmologies. It is obvious from \eqref{Cs}, \eqref{einstein2} and \eqref{conditionX} that constant scalar curvature is possible only if $R=0$, which corresponds to either $Y={\rm const.}$ (see the previous paragraph) or $F_{mn}F^{mn}=0$. Hence, de Sitter solution ($R=4\Lambda$) is not possible in our model, but inflationary cosmologies via varying scalar curvature (e.g., power-law inflation) can be studied. This implies that broken scale invariance, which is confirmed by observations, is a typical feature of an inflationary Universe obtained in our model. Another interesting point that can be seen from the constraint \eqref{conditionX} that characterises our model is that because $F_{mn}F^{mn}$ is positive definite, $\frac{{\rm d} Y}{{\rm d} R}<0$ for $R>0$. This differentiates our model from the inflationary cosmologies based on the non-minimal model of gravity in $\frac{R^\eta}{M^{2\eta}}F^2$-form in the literature \cite{Turner:1987bw,Mazzitelli:1995mp,Campanelli:2008qp,Lambiase:2008zz}, even if we would consider $Y=\frac{R^\eta}{M^{2\eta}}$ as an additional constraint, i.e., substitute $Y=\frac{R^\eta}{M^{2\eta}}$ in \eqref{conditionX}. Namely, if we assume $Y=\frac{R}{M^2}$, where $M$ is some mass scale, then we would get $F_{mn}F^{mn}=-\frac{2M^2}{\kappa^2}$, which is not allowed since $F_{mn}F^{mn}$ is positive definite, and hence $Y=\frac{R}{M^2}$ is forbidden in our model. More generally, if we assume $Y=\frac{R^\eta}{M^{2\eta}}$, then we would find from the constraint \eqref{conditionX} that $\frac{\eta}{M^{2\eta}} R^{\eta-1}  F_{mn}F^{mn}=-\frac{2}{\kappa^2}$, which implies that $\eta$ should be negative as long as $R>0$. However, the models studied in \cite{Turner:1987bw,Mazzitelli:1995mp,Campanelli:2008qp,Lambiase:2008zz} assume that $\eta$ is a positive integer, whereas our model demands negative $\eta$ values that are not necessarily integers. Let us next show that this consequent model has indeed features that make it worthy of further investigation in the context of anisotropic inflationary models that can violate the cosmic no-hair conjecture. In what follows, we continue with a brief discussion on two possible interpretations of this model that would be useful in this regard.

One way of looking at this system of field equations is to consider it as the trace-free Einstein gravity \cite{Einstein:1919gv,Weinberg:1988cp}, which has been recently studied under the name of unimodular gravity \cite{Unruh:1988in,Ellis:2010uc}, coupled to Maxwell EMT with a variable ($R$-dependent) coupling parameter. That is, from \eqref{einstein2}, we have
\begin{eqnarray}
\label{einstein5} 
* R_a - \frac{1}{4} R*e_a  =   \frac{\kappa^2}{4} Y(\iota_a F \w *F - F \w \iota_a *F),
\end{eqnarray}
together with the constraint \eqref{conditionX} and the modified Maxwell equations \eqref{maxwell1}. This shows that the model is able to yield accelerated expansion, since it is a feature of unimodular gravity that the cosmological constant appears as an integration constant. The $R$ dependent variable coupling due to the $Y(R)$ term, on the other hand, can permit anisotropic stress of the electromagnetic field, rather than a cosmological constant, to be effective in controlling the expansion anisotropy provided that a suitable function for $Y(R)$ is chosen.

 
In fact, alternatively, we can stick to the conventional GR, which also gives us the opportunity to further investigate the features of the model that is determined by some interesting properties of the effective EMT. To do so, adding the Einstein tensor $G_a$ to both sides of \eqref{einstein2} and then taking one half of the resultant equation, we rewrite \eqref{einstein2} as follows;
\begin{eqnarray}
\label{eqn:forbbn}
*R_a - \frac{1}{2} R  *e_a    = \kappa^2 \frac{Y}{4} (\iota_a F \w *F - F \w \iota_a *F) - \frac{1}{4} R  *e_a.
 \end{eqnarray}
This recasts the effective EMT to
 \begin{equation}
\label{effemtf}
 \tau'_a=\frac{1}{4} Y(\iota_a F \w *F - F \w \iota_a *F) - \frac{1}{4\kappa^2}R  *e_a
 \end{equation}
subject to the constraint \eqref{conditionX} and the modified Maxwell equations \eqref{maxwell1}. We note that the last term of  $\tau'_a$ \eqref{effemtf} yields the form of the conventional vacuum EMT defined as $\tau_a^{(\rm vac)}=-\rho^{(\rm vac)} *e_a$. Accordingly, \eqref{effemtf} can be decomposed into two additive components as follows:
 \begin{equation}
 \label{eqn:taup}
 \tau'_a= \tau_a^{(\rm mEMF)}+ \tau_a^{(\rm vac)} ,
  \end{equation}
where
\begin{equation}
\label{eqn:vac}
\tau_a^{(\rm vac)}=-\rho^{(\rm vac)} *e_a=- \frac{R}{4\kappa^2}  *e_a
 \end{equation}
is the vacuum energy EMT with a certain energy density $\rho^{(\rm vac)}=\frac{R}{4\kappa^2}$, and
\begin{equation}
\label{eqn:mem}
\tau_a^{(\rm mEMF)}=\frac{Y}{2}\, \tau_a^{({\rm em})}=\frac{Y}{2}\frac{1}{2}(\iota_a F \w *F - F \w \iota_a *F)
\end{equation}
is the EMT of the modified electromagnetic field (mEMF), such that the conventional electromagnetic field EMT is rescaled with a dynamical factor $Y/2$. Here $\tau_a^{(\rm vac)}$, which yields the equation of state $p^{(\rm vac)}=-\rho^{(\rm vac)}$, is the component that leads to accelerated expansion, while $\tau_a^{(\rm mEMF)}$, which yields anisotropic pressure in the form $[p_x,p_y,p_z]=[-\rho,\rho,\rho]$, is the component that adjusts the isotropisation imposed by the vacuum energy. As dictated by the cosmic no-hair theorem, an initially expanding spatially homogeneous general relativistic cosmological model in the presence of a conventional electromagnetic field and vacuum energy would evolve towards the de Sitter solution on an exponentially rapid time scale \cite{Wald:1983ky}. In our case, on the other hand, the EMTs of the mEMF $\tau_a^{(\rm mEMF)}$ and the vacuum energy $ \tau_a^{(\rm vac)}$ are non-minimally coupled through a dynamical factor $Y/2$, and hence the modified electromagnetic field can be persistent by transferring energy-momentum from the vacuum energy so that its anisotropic pressure can still be effective and can even maintain an anisotropic expansion. Therefore, taking the covariant derivative of the two components and representing the interaction between them by $Q$, we can write
\begin{equation}
\label{eqn:trans1}
{\rm D}\tau_a^{(\rm mEMF)}=Q*e_a \quad\textnormal{and}\quad {\rm D}\tau_a^{(\rm vac)}=-Q*e_a.
\end{equation}
Using \eqref{eqn:vac} in \eqref{eqn:trans1} we find $Q=-\dot{\rho}^{({\rm vac})}=-\frac{\dot{R}}{4\kappa^2}$ which together with the constraint equation \eqref{conditionX} leads to
\begin{equation}
\label{eqn:trans2}
Q=\frac{F_{mn}F^{mn}}{8}\dot{Y}.
\end{equation}
We see from the above discussions that de Sitter inflation ($R=4\Lambda$) is not allowed in our model, but inflationary cosmologies with $R>0$ and $\dot{R}\lesssim 0$ (e.g., power-law inflation) can be studied. In accordance with that, because $F_{mn}F^{mn}>0$, it is easy to see from the constraint \eqref{conditionX} that $\dot{R}<0$ leads to $\dot{Y}>0$, which in turn leads to $Q>0$ as can be seen from \eqref{eqn:trans2}. It follows from \eqref{eqn:trans1} and \eqref{eqn:trans2} that the mEMF $\tau_a^{(\rm mEMF)}$ transfers energy-momentum from vacuum energy $ \tau_a^{(\rm vac)}$. Hence, it could be possible to avoid the mEMF to be wiped out by the vacuum energy, which in turn could avoid the rapid isotropisation of the Universe as it expands and give us opportunity to keep the isotropisation under control. In addition, it could be possible that $Y$ increases rapidly enough to suspend or even increase the energy density of the mEMF during the inflationary era at/to a value that might be of astrophysical interest in the post-inflationary Universe, e.g., at/to a value that might account for the presence of large-scale cosmological magnetic fields we observe today. Finally, we see that the an/isotropisation of the Universe during inflation, the deviation from the temporal de Sitter symmetry (which is inevitable in our model) and the generation/persistence of the mEMF are strongly intertwined with each other in a particular way characterised by the constraint \eqref{conditionX}. Namely, the larger the deviation from the temporal de Sitter symmetry ($\dot{R}=0$) the larger the generation/stronger the persistence of the electromagnetic field, which in turn implies that a stronger effect of the anisotropic pressure of the electromagnetic field on the evolution of the anisotropy than that of the vacuum energy (the best isotropizer \cite{Starobinsky:1982mr}). Indeed, in what follows we construct a cosmological model satisfying the constraint \eqref{conditionX} and show first that the widely studied $\frac{R^\eta}{M^{2\eta}}F^2$-form coupling \cite{Horndeski:1976gi,Turner:1987bw,MuellerHoissen:1988bp,Balakin:2010ar,Dereli:2011hu,Baykal:2015paa,Mazzitelli:1995mp,Bamba:2008ja,Campanelli:2008qp,Lambiase:2008zz,Kunze:2009bs,Dereli:2011gh} appears as the case that sets the ratio of the energy densities of the mEMF and vacuum $\rho^{({\rm mEMF)}}/\rho^{({\rm vac)}}$ to a constant depending on $\eta$, and then we further study an inflationary cosmology under the power-law volumetric assumption, which gives $\frac{R^\eta}{M^{2\eta}}F^2$-form coupling just as a special case.


 \section{Cosmological model}
 \label{sec:cosmo}

It is shown above that the theory under consideration is equivalent to GR
\begin{equation}
\label{GR}
*R_a - \frac{1}{2} R  *e_a=\kappa^2 \tau'_a
\end{equation}
with the following effective EMT, which is in fact the total energy density of the non-minimally interacting mEMF and vacuum \eqref{eqn:taup},
\begin{equation}\label{semt}
 \tau'_a=\frac{1}{4} Y(\iota_a F \w *F - F \w \iota_a *F) - \frac{1}{4\kappa^2}R  *e_a \ , 
\end{equation}
plus the modified Maxwell equations and an additional differential constraint
\begin{equation}
\label{sbjt}
{\rm d}F=0 \ ,\quad {\rm d}(Y * F) = 0  \quad \textnormal{and}\quad Y_RF_{mn}F^{mn} = -\frac{2}{\kappa^2} .
\end{equation}
In what follows, we will investigate a cosmological model based on these equations.

We assume an electromagnetic field 2-form with a single magnetic component $B=B(t)$ along the $x$-direction
 \begin{equation}\label{electromagnetic1}
 F   = B e^{23} \ ,
 \end{equation}
and consider the simplest spacetime metric that can accommodate this choice. Namely, the spatially flat and homogeneous but not necessarily isotropic locally rotationally symmetric (LRS) Bianchi type-I spacetime metric,
\begin{equation}\label{metric}
              g = - {\rm d} t^2  +  a^{2} {\rm d} x^2 + b^2 ( {\rm d} y^2 + {\rm d} z^2).
\end{equation}
Here  $a=a(t)$ and $b=b(t)$ are directional scale factors along the $x$-axis and the $y$- and $z$-axes, respectively, and are functions of the cosmic time $t$ only. Thus the orthonormal co-frame 1-forms read
\begin{eqnarray}
e^0 = {\rm d}t, \quad e^1 = a {\rm d}x ,   \quad e^2 = b {\rm d}y , \quad e^3 = b {\rm d}z.
\end{eqnarray}
The $\lambda _a $ variations of (\ref{lag1}) impose the zero-torsion constraint
\begin{eqnarray}
{\rm d} e^a +{\omega^a}_b \wedge e^b=0 \ , 
\end{eqnarray}  
and we calculate the non-zero Levi-Civita connection 1-forms as
\begin{eqnarray}
\omega^{01} = \frac{\dot{a}}{a} e^1 \ ,  \quad \omega^{02} = \frac{\dot{b}}{b} e^2 \ , \quad \omega^{03} = \frac{\dot{b}}{b} e^3 \ ,
\end{eqnarray}
where a dot denotes a derivative with respect to the cosmic time $t$. It follows that the corresponding non-zero components of the curvature 2-forms are
\begin{eqnarray}
R^{01} = \frac{\ddot{a}}{a} e^{01} , \quad
R^{02} = \frac{\ddot{b}}{b} e^{02} , \quad
R^{03} = \frac{\ddot{b} }{b} e^{03} , \quad
R^{12} = \frac{\dot{a} \dot{b} }{a b} e^{12} , \quad
R^{13} = \frac{\dot{a} \dot{b} }{a b} e^{13}, \quad
R^{23} = \frac{ \dot{b}^2 }{b^2} e^{13} . 
\end{eqnarray}
Then the Ricci curvature 1-forms are found to be
\begin{align}
R_0 = - \left(\frac{\ddot{a}}{a}  +  2 \frac{\ddot{b}}{b} \right)e^0, \quad  R_1 =  \left(\frac{\ddot{a}}{a}  +  2 \frac{\dot{a}\dot{b} }{a b} \right)e^1,\quad
 R_2 =  \left(\frac{\ddot{b}}{b}  +   \frac{\dot{a}\dot{b} }{a b} + \frac{{\dot{b}}^2}{b^2} \right)e^2, \quad  R_3 = \left( \frac{\ddot{b}}{b}  +   \frac{\dot{a}\dot{b} }{a b} + \frac{{\dot{b}}^2}{b^2} \right) e^3.
\end{align}
Finally the curvature scalar turns out to be
\begin{eqnarray}\label{R}
R = 2 \frac{\ddot{a}}{a} + 4\frac{\ddot{b}}{b} + 4\frac{\dot{a}\dot{b} }{a b} + 2\frac{{\dot{b}}^2}{b^2}\ .
\end{eqnarray}

Using all these results in \eqref{GR}-\eqref{semt}, we reach the following non-linear system of second order differential equations:
\begin{eqnarray}
2\frac{\dot{a}\dot{b} }{a b} + \frac{{\dot{b}}^2}{b^2} & =&  \kappa^2 \frac{YB^2}{4} +\frac{1}{2} \frac{\ddot{a}}{a} + \frac{\ddot{b}}{b} + \frac{\dot{a}\dot{b} }{a b} + \frac{1}{2}\frac{{\dot{b}}^2}{b^2},
\label{22}\\
2\frac{\ddot{b} }{ b} +  \frac{{\dot{b}}^2}{b^2}  &=&  \kappa^2 \frac{YB^2}{4} +\frac{1}{2} \frac{\ddot{a}}{a} + \frac{\ddot{b}}{b} + \frac{\dot{a}\dot{b} }{a b} + \frac{1}{2}\frac{{\dot{b}}^2}{b^2}, \label{23} 
 \\
\frac{\ddot{a} }{ a}   + \frac{\ddot{b} }{ b} +  \frac{\dot{a}\dot{b} }{a b}&=& -\kappa^2 \frac{YB^2}{4} +\frac{1}{2} \frac{\ddot{a}}{a} + \frac{\ddot{b}}{b} + \frac{\dot{a}\dot{b} }{a b} + \frac{1}{2}\frac{{\dot{b}}^2}{b^2},
\label{24}
\end{eqnarray}
subject to the constraint
\begin{equation}
\label{concon}
 B^2 \frac{{\rm d}Y}{{\rm d} R}=-\frac{2}{\kappa^2}.
\end{equation}
We must set
\begin{eqnarray}
\label{B}
B=\frac{B_0}{b^2}
\end{eqnarray}
with an integration constant  $B_0$ to satisfy the homogeneous Maxwell equation ${\rm d}F=0$ in \eqref{sbjt}. We note that this magnetic field satisfies the inhomogeneous Maxwell equation $ {\rm d}(Y * F) = 0 $
in \eqref{sbjt} automatically. We next note that any one of equations in \eqref{22}-\eqref{24} can always be written in terms of other two and hence we have in fact only two linearly independent equations. Thus it is enough for us to consider only two of these three equations. Now subtracting \eqref{23} from \eqref{22} we immediately find
\begin{eqnarray}\label{40}
\frac{\ddot{b}}{b}  = \frac{\dot{a}\dot{b}}{ab}.
\end{eqnarray}
Its solution leads to the following relation between the directional scale factors
\begin{eqnarray}\label{a}
 a = t_0\dot{b},
\end{eqnarray}
where $t_0$ is an integration constant. Next using \eqref{a} as well as $Y$ from \eqref{22} (or \eqref{23}) and $R$ from \eqref{R}, $B$ from \eqref{B} we find that the system of equations \eqref{22}-\eqref{23} always satisfy the condition \eqref{concon}. Hence any solution we obtain for \eqref{22}-\eqref{23} would automatically satisfy \eqref{concon}. Finally using \eqref{40} and \eqref{B} either in \eqref{22} or in \eqref{23}, we reach the following equation for $Y$:
\begin{equation}
\label{neq}
Y=\frac{2b^4}{\kappa^2 \ B_0^2}\left( \frac{\dot{b}^2}{b^2}-\frac{\ddot{a}}{a} \right).
\end{equation}
Hence the final set of equations that would describe our model would be given by \eqref{a} and \eqref{neq} to be satisfied by three unknown functions $a$, $b$ and $Y$. Therefore, in contrast to a general relativistic cosmological model in the presence of minimally coupled conventional electromagnetic field solely, our model is not fully determined and we have a freedom to introduce one more equation that might lead to interesting cosmological models. Accordingly, in Section \ref{sec:plve}, we shall investigate an inflationary cosmology under the {\it power-law volumetric expansion} assumption. However, it would be convenient to introduce some cosmological parameters that we shall later use and discuss some general properties of such a cosmological model.

\subsection{Some general properties of the model}
\label{sec:gen}

Let us start by the properties of the effective fluid described in \eqref{eqn:taup} by the EMT $\tau'_a=T_{ab} * e^b$ which gives
\begin{equation}
T_{ab} = {\rm diag}[\rho, p_x,p_y,p_z],
\end{equation}
where $\rho$ is the energy density and $p_x$ and $p_y=p_z\equiv p_{y,z}$ are the directional pressures along the $x$-axis and the $y$- and $z$-axes, respectively. Accordingly the energy density of this effective fluid that appears in \eqref{22} reads
\begin{equation}
\label{eqn:rhocos}
\rho= \frac{YB^2}{4}+\frac{R}{4\kappa^2},
\end{equation}
which is in fact the total energy density of the non-minimally interacting mEMF and vacuum, whose energy densities are given by
\begin{equation}
\label{eqn:memvac}
\rho^{(\rm mEMF)}=\frac{YB^2}{4}\quad\textnormal{and}\quad \rho^{(\rm vac)}=\frac{R}{4\kappa^2},
\end{equation}
as can be seen from \eqref{eqn:vac} and  \eqref{eqn:mem}.

The directional pressures that appear in \eqref{23} and \eqref{24} read
\begin{equation}
p_x= - \ \frac{YB^2}{4} -\frac{R}{4\kappa^2}\quad\textnormal{and}\quad 
p_{y,z} =   \frac{YB^2}{4} -\frac{R}{4\kappa^2},
\end{equation}
that lead to the following directional equation of state parameters;
\begin{equation}
\label{finaleos}
w_x=\frac{p_x}{\rho}  =  - 1
\quad \textnormal{and}\quad
w_{y,z}=\frac{p_{y,z}}{\rho}=  \frac{YB^2- \frac{R}{\kappa^2 }}{YB^2+ \frac{R}{\kappa^2}}.
\end{equation}
Thus we have an effective fluid described by the above dynamically anisotropic EoS \eqref{finaleos} that precisely yields $-1$ along the $x$-axis as that of the vacuum energy and, provided that $Y\geq 0$, the form of canonical scalar fields' EoS parameter and ranges in the region $-1\leq w_{y,z} \leq 1$ along the $y$- and $z$-axes. Accordingly, (i) if $YB^2\ll \frac{R}{\kappa^2}$ then the EoS parameter becomes $w_x=w_{y,z}=-1$, i.e. the fluid mimics vacuum energy, (ii) if $YB^2\gg \frac{R}{\kappa^2 }$ then the EoS parameter becomes $[w_x,w_y,w_z]=[-1,1,1]$, i.e. the fluid mimics the conventional electromagnetic field, and (iii) if $YB^2= \frac{R}{\kappa^2 }$ then the EoS parameter becomes $[w_x,w_y,w_z]=[-1,0,0]$, i.e. the fluid mimics cosmic string along the $x$-axis. Moreover, we note that if $Y<0$, then the energy density of the effective fluid could still be positive $\rho>0$ as it should be, yet now the EoS along the $y$- and $z$-axes would cross the phantom divide line $w_{y,z}<-1$ implying that the energy density would increase as the Universe expands like in case of phantom fields. Therefore the study of the dynamics of the Universe driven by such a fluid would be interesting since it not only leads to accelerating expansion but also modifies the expansion anisotropy.

We can show at this point that the an/isotropisation of the Universe during inflation, the deviation from the temporal de Sitter symmetry, and the generation/persistence of the mEMF are strongly intertwined with each other in a particular way characterised by the contraint \eqref{conditionX}.  We find, using \eqref{eqn:rhocos} and \eqref{eqn:memvac} in \eqref{concon}, that the ratio of the energy density of the mEMF to the total energy density is
\begin{equation}
\frac{\rho^{(\rm mEMF)}}{\rho}=\left[1-\frac{R}{2Y}\frac{{ \rm d} Y}{{\rm d} R}\right]^{-1}.
\end{equation}
We see from this equation that $\frac{\rho^{(\rm mEMF)}}{\rho}=\frac{2}{2-\eta}$ for $Y\propto R^\eta$ with $\eta<0$, namely, the mEMF does not dilute as the Universe expands, but becomes persistent with respect to the vacuum energy. According to this the non-minimal model of gravity in $\frac{R^\eta}{M^{2\eta}}F^2$-form corresponds to a particular case of our model that renders the ratio of the energy density of the mEMF to the total energy density constant. It is noteworthy that because the ratio $\frac{\rho^{(\rm mEMF)}}{\rho}$ would obviously determine the deviation from the temporal de Sitter symmetry and hence the deviation from scale invariant power spectrum as well, it is possible to write the scalar spectral index $n_s$, which measures the slight deviation from scale invariance, as a specific function of $\eta$. In addition, we find from \eqref{finaleos} that $\eta$ also determines the total anisotropic EoS as $[w_x,w_y,w_z]=[-1,-1+\frac{4}{2-\eta},-1+\frac{4}{2-\eta}]$, which in turn determines the expansion anisotropy. We would also like to note here that non-power-law functions of $Y(R)$ would lead to varying $\frac{\rho^{(\rm mEMF)}}{\rho}$, which in turn can lead to a non-trivial an/isotropization during inflation and, for instance, the amplification of the mEMF against to the vacuum energy during inflation. Indeed, in the next section, we give an exact solution of the model under the power-law volumetric assumption and show that $Y\propto R^\eta$-form coupling appears just as a special case of the solution.

We note that \eqref{a} holds a central place in the characterisation of our model; any fully determined model that might be obtained by introducing an additional constraint equation should obey this relation. Then just looking at this relation, one may immediately see two noteworthy properties of the model: (i) The same rate of expansion along the all directions $\dot{a}=\dot{b}$, i.e. isotropic expansion, is possible only in case of all axes expanding exponentially, and any deviation from exponential expansion results in a deviation from isotropic expansion. (ii) The expansion of the $x$-axis, $\dot{a}>0$, independent of whether the $x$-axis exhibits accelerated or decelerated expansion, accompanied by an accelerated expansion of the $y$- and $z$-axes $\dot{b}>0$ and $\ddot{b}>0$. For a further discussion, it would be convenient to introduce here some parameters that are of interest in cosmology. We define the mean scale factor $v$, the average Hubble parameter $H$ and the deceleration parameter of the volumetric expansion $q$, respectively, as follows:
\begin{equation}
v=\left (ab^2 \right )^{\frac{1}{3}},\quad H=\frac{\dot{v}}{v}=\frac{1}{3}\left(\frac{\dot{a}}{a}+2\frac{\dot{b}}{b}\right), \quad q= -\frac{v\ddot{v}}{\dot{v}^2}= -1 + \frac{{\rm d}}{{\rm d}t}\left(\frac{1}{H}\right).
\end{equation}
In a similar way, the directional Hubble and deceleration parameters along the $x$-, $y$- and $z$-axes are given as follows:
\begin{equation}
H_{x}=\frac{\dot{a}}{a},\quad H_{y,z}=\frac{\dot{b}}{b}, \quad  q_{x}=-1+\frac{{\rm d}}{{\rm d}t}\left(\frac{1}{H_{x}}\right),\quad q_{y,z}=-1+\frac{{\rm d}}{{\rm d}t}\left(\frac{1}{H_{z}}\right).
\end{equation}
The deceleration parameters can be taken as the key parameters among the others, because for any expanding scale factor (namely $v$, $a$ or $b$), the negative values of the corresponding deceleration parameter imply acceleration, positive values imply deceleration and the special values $-1$ and $0$ correspond to exponential expansion and constant-rate expansion, respectively. Two other parameters relevant to the discussion of anisotropic cosmological models are the shear scalar $\sigma^{2}$ and the dimensionless expansion anisotropy parameter $\Delta$ that are defined as follows:
\begin{equation}
\label{eqn:aniso1}
\sigma^{2}=\frac{1}{2}\sum_{i=1}^{3}\left(H_{i}-H\right)^{2} \quad\textnormal{and}\quad \Delta=\frac{1}{3}\sum_{i=1}^{3}\left(\frac{H_{i}-H}{H}\right)^{2}.
\end{equation}
Here the sums are on  $(1,2,3)=(x,y,z)$. $\Delta$ is a measure of the deviation from isotropic expansion; the Universe expands isotropically for $\Delta=0$, and its time-evolution is crucial for deciding whether the expansion of the space leads to the isotropization of the geometry of the space. Namely, the spatial section of the metric approaches isotropy for $v\rightarrow\infty$ and $\Delta\rightarrow 0$ as $t\rightarrow \infty$ \cite{Collins:1972tf}. It is also important to note that $\Delta$ cannot take arbitrary values but ranges between $0\leq \Delta \leq 2$, as long as the average energy density of the Universe is non-negative due to the relation $3(1-\Delta /2)H^2=\kappa^2\rho$ for Bianchi type-I spacetimes in GR \cite{Akarsu:2010zm}.  Subtracting \eqref{24} from \eqref{23} we find 
\begin{eqnarray}
\frac{\rm d}{{\rm d}t}(H_y -H_x)  +3H(H_y - H_x)= \frac{\kappa^2YB^2}{2}.
\end{eqnarray}
Its integration gives the difference between the directional Hubble parameters along the  $x$-axis and  $y$- and $z$-axes as
 \begin{eqnarray}
 H_y -H_x = \frac{\lambda}{ 2 V} +   \frac{\kappa^2}{ 2 V} \int{YB^2V {\rm d}t},
 \end{eqnarray}
where $V=v^3$ is the volume scale factor. Accordingly we find that the shear scalar and expansion anisotropy parameter are as follows:
\begin{eqnarray}
\sigma^2 = \frac{1}{3} \left[      \frac{\lambda}{ 2 V}  +    \frac{\kappa^2}{ 2 V} \int{YB^2V {\rm d}t} \right ]^2 \quad\textnormal{and}\quad \Delta = \frac{2}{9H^2} \left[      \frac{\lambda}{ 2 V}  +    \frac{\kappa^2}{ 2 V} \int{YB^2V {\rm d}t} \right ]^2.
\end{eqnarray}
We note that the presence of a non-minimal coupling, i.e. the function $Y$ appearing under the integrals above, has a crucial role in the determination of the evolution of the expansion anisotropy. For a comparison, one may check that, in GR, if a Bianchi type-I Universe is filled with isotropic sources only then the shear scalar and expansion anisotropy parameter would evolve simply as $\sigma^2\propto V^{-2}$ and $\Delta\propto \dot{V}^{-2} $ \cite{Akarsu:2010zm,Barrow:1995fn,Ellis:1998ct}.

\section{Anisotropic inflation with power-law volumetric expansion}
\label{sec:plve}
In this study we consider neither a scalar field nor a RW spacetime, but an anisotropic fluid and spacetime, and we shall focus on the properties of the expansion of the Universe during inflation. Hence it would be more convenient to carry out our discussions by considering the Hubble-slow-roll approximation rather than the potential-slow-roll approximation that relies on the presence of a scalar field/s \cite{Copeland:1993jj,Liddle:1994dx}. Accordingly, we would say that inflationary models are mostly characterised by a Friedmann-Robertson-Walker (FRW) Universe that exhibits quasi de Sitter (exponential) expansion. The power-law inflation \cite{Lucchin:1984yf} models provide the simplest generalisation of de Sitter inflation models. In a spatially flat FRW Universe the power-law expansion corresponds to $V\propto t^{\frac{3}{\epsilon}}$ where $\epsilon=q+1$ is the positive definite slow-roll parameter for which $\epsilon<1$ corresponds to the power-law inflation, namely, $\epsilon\lesssim 0.05$ for the near-de Sitter inflation and $\epsilon\approx 0$ for the almost-de Sitter inflation. We assume in this study that the {\it volume} of the spatially flat but not necessarily isotropic Universe we consider exhibits a power-law inflation:
\begin{equation}
\label{plvol}
V=ab^2=V_1 t^{3\alpha-1},
\end{equation}
which leads to the following mean scale factor, Hubble and deceleration parameters  
\begin{equation}
v=v_1 t^{\alpha-\frac{1}{3}}, \quad H=\frac{3\alpha-1}{3t}\quad\textnormal{and}\quad q=\frac{3}{3\alpha-1}-1,
\end{equation}
where $\alpha \equiv \frac{1}{\epsilon}+\frac{1}{3}$. The power-law expansion of the mean scale factor {\it does not necessarily} lead to directional scale factors exhibiting power-law expansion: Using \eqref{a} with the power-law volumetric expansion assumption \eqref{plvol} we find the kinematics of the $x$-axis as 
\begin{equation}
\begin{aligned}
a=v_1 \alpha^{\frac{2}{3}} t_0^{\frac{2}{3}} t^{3\alpha-1} (t^{3\alpha}+k)^{-\frac{2}{3}}, \quad H_x=\frac{(3\alpha-1)k+(\alpha-1)t^{3\alpha}}{(t^{3\alpha}+k)t} \quad\textnormal{and}\quad\\
q_x=\frac{2k\alpha(3\alpha-1) \left[(\alpha-1)t^{3\alpha}-k\right] }{(\alpha-1)\left[(\alpha-1)t^{3\alpha}+(3\alpha-1)k\right]^{2}}+\frac{1}{\alpha-1}-1,
\end{aligned}
\end{equation}
and the kinematics of the $y$- and $z$-axes as
\begin{equation}
b=\frac{v_1}{\alpha^{\frac{1}{3}} t_0^{\frac{1}{3}}} (t^{3\alpha}+k)^{\frac{1}{3}}, \quad H_{y,z}=\frac{\alpha t^{3\alpha}}{(t^{3\alpha}+k)t} \quad\textnormal{and}\quad q_{y,z}=\frac{-k(3\alpha-1)t^{-3\alpha}}{\alpha}+\frac{1}{\alpha}-1.
\end{equation}
Here $k\geq 0$ is integration constant. It follows that the shear scalar and expansion anisotropy read
\begin{eqnarray}
\label{ani}
\sigma^2=\frac{(t^{3\alpha}-3k\alpha+k)^2}{3(t^{3\alpha}+k)^2t^2}\quad\textnormal{and}\quad \Delta = \frac{2(t^{3\alpha}-3k\alpha+k)^2}{(t^{3\alpha}+k)^2(3\alpha -1)^2}.
\end{eqnarray}
In the next step, using the scale factors, we find that the energy density and the directional pressures of the effective fluid read
\begin{equation}
\begin{aligned}
\rho&=\left[1+\frac{3\alpha k}{3\alpha-2}(t^{3\alpha}+k)^{-1}\right]\frac{t^{3\alpha}}{t^{3\alpha}+k}\frac{\alpha(3\alpha-2)}{\kappa^2}\,t^{-2}\\
p_x   = -\rho,\quad p_{y,z} &= -\frac{2k(\alpha-2)(3\alpha-1)t^{3\alpha}+(3\alpha-1)(3\alpha-2)k^2 + (3\alpha-2)(\alpha-1) t^{6\alpha}   }{\kappa^2 (t^{3\alpha}+k)^2 t^2};
\end{aligned}
\end{equation}
thus leading to the following directional EoS parameters
\begin{eqnarray}
w_x = -1 \quad\textnormal{and}\quad w_{y,z} = -\frac{(t^{3\alpha}+k)[2k(\alpha-2)(3\alpha-1)t^{3\alpha}+(3\alpha-1)(3\alpha-2)k^2 + (3\alpha-2)(\alpha-1) t^{6\alpha}] }{2\alpha(3\alpha-1)k^2 t^{3\alpha}+\alpha(9\alpha-4)k t^{6\alpha}+\alpha(3\alpha-2)t^{9\alpha}}.
\end{eqnarray}
Substituting the directional scale factors we give above in \eqref{neq}, we find that the function that determines our non-minimal coupling is
\begin{equation}
\label{yt}
  Y = \frac{ 2v_1^4  \left[(3\alpha-2)t^{6\alpha}  +4k(3\alpha-1)t^{3\alpha} - k^2 (3\alpha-1)(3\alpha-2)\right]}{\kappa^2 B_0^2 \alpha^{\frac{4}{3}} t_0^{\frac{4}{3}}(t^{3\alpha}+k)^{\frac{2}{3}} t^2}.
\end{equation}
Finally, using this equation \eqref{yt} and \eqref{B} in \eqref{eqn:memvac}, we obtain the energy density of the mEMF as
\begin{align}
\label{eqn:Gmemvac}
\rho^{(\rm mEMF)} =\frac{3\alpha -2  }{2\kappa^2 }\, t^{-2} + \frac{3\alpha k}{\kappa^2}\left[(t^{3\alpha}+k)^{-1}- \frac{3\alpha k}{2} (t^{3\alpha}+k)^{-2}\right]\, t^{-2}.
\end{align}

We observe from the general solution, obtained under the {\it power-law volumetric expansion} assumption, that the directional scale factors do not exhibit simple power-law expansion unless $k=0$. We note on the other hand that the model approaches the particular case $k=0$ for large $t$ values, namely, when $t^{3\alpha}\gg k$. Therefore giving the particular case $k=0$ explicitly would be useful to show where the Universe in our model will eventually evolve into. Accordingly we look at the particular case $k=0$, for which not only the mean scale factor but also  the directional scale factors exhibit power-law expansion:
\begin{equation}
\label{kinx}
a=v_1 \alpha^{\frac{2}{3}} t_0^{\frac{2}{3}} t^{\alpha-1}, \quad H_x=\frac{\alpha-1}{t} \quad\textnormal{and}\quad q_x=-1+\frac{1}{\alpha-1},
\end{equation}
and
\begin{equation}
\label{kiny}
b=\frac{v_1}{\alpha^{\frac{1}{3}} t_0^{\frac{1}{3}}} t^{\alpha}, \quad H_{y,z}=\frac{\alpha}{t} \quad\textnormal{and}\quad q_{y,z}=-1+\frac{1}{\alpha}.
\end{equation}
The scalar curvature for this case becomes
\begin{equation} 
\label{eqn:Rsimple}
R=2(3\alpha-2)(2\alpha-1)\,t^{-2}.
\end{equation}
The shear scalar simplifies so that it is now inversely proportional to the square of cosmic time $t$ while the expansion anisotropy is a constant:
\begin{eqnarray}
\label{anik0}
\sigma^2=\frac{1}{3t^2}\propto V^{-\frac{2}{3\alpha-1}}\quad\textnormal{and}\quad \Delta = \frac{2}{(3\alpha -1)^2}.
\end{eqnarray}
Using \eqref{kinx}, \eqref{kiny} and \eqref{anik0}, we derive the following interesting relation between the expansion anisotropy and the average deceleration parameter:
\begin{eqnarray}
  \Delta = \frac{2}{9}(q+1)^2=\frac{2}{9}\epsilon^2.
\end{eqnarray}
The energy density and directional pressures of the effective fluid now read
\begin{eqnarray}
\label{eqn:rhoSolr}
\rho=  \frac {\alpha\, \left( 3\, \alpha-2 \right) }{  \kappa^2 } t^{-2}, \quad
p_x   = -\rho,\quad p_{y,z} = -{\frac { \left( 3\,\alpha-2 \right)  \left( \alpha-1 \right) }{\kappa^2 }  } t^{-2},
\end{eqnarray}
and imply the directional EoS parameters
\begin{eqnarray}
w_x = -1 \quad\textnormal{and}\quad w_{y,z} = -1 +\frac{1}{\alpha}.
\end{eqnarray}
We note that in this particular case not only the directional EoS parameter along the $x$-axis is a constant but also the EoS parameters along the $y$- and $z$-axes, all with a dependence on the parameter $\alpha$.
From \eqref{24} we see that, in the special case $k=0$, we have
\begin{equation}
 Y =   \frac{2 v_1^4  ( 3\alpha -2  ) }{\kappa^2 B_0^2 t_0^{\frac{4}{3}} \alpha^{\frac{4}{3}} } \, t ^{4\alpha -2 } \propto V^{\frac{4\alpha-2}{3\alpha-1}}.
\end{equation}
Then we may solve for the function $Y$ in terms of the curvature scalar $R$ \eqref{eqn:Rsimple} for the particular case $k=0$ as
\begin{equation}
\label{YBRcheck}
Y(R) = \frac{ v_1^4 \ 2^{2\alpha} \ (3\alpha -2)^{2\alpha}   \ (2\alpha -1 )^{2\alpha-1}  }{ \alpha^{\frac{4}{3}} \ t_0^{\frac{4}{3}} \ \kappa^2 \ B_0^2 }  \, R^{1-2\alpha},
\end{equation}
which has the form of one of the most widely considered generalisation of Einstein-Maxwell theory (see e.g. \cite{Horndeski:1976gi,Turner:1987bw,MuellerHoissen:1988bp,Balakin:2010ar,Dereli:2011hu,Baykal:2015paa,Mazzitelli:1995mp,Bamba:2008ja,Campanelli:2008qp,Lambiase:2008zz,Kunze:2009bs,Dereli:2011gh}) especially for the generation of seed magnetic fields during the inflation \cite{Turner:1987bw,Mazzitelli:1995mp,Campanelli:2008qp,Lambiase:2008zz}. We obtain the energy density of the mEMF as
\begin{equation}
\label{eqn:memr}
\rho^{(\rm mEMF)}=\frac{3\alpha -2  }{2\kappa^2 }\, t^{-2},
\end{equation}
which can also be written in terms of scale factor along the $y$-axis as
\begin{equation}
\rho^{(\rm mEMF)}=\frac{3\alpha -2  }{2\kappa^2 }\left(\frac{\alpha^{\frac{1}{3}}t_0^{\frac{1}{3}}}{v_1}\right)^{-\frac{2}{\alpha}}\, b^{-\frac{2}{\alpha}}.
\end{equation}
It is noteworthy that, for $\alpha>\frac{1}{2}$, the energy density of the mEMF decreases slower than that of the standard electromagnetic field, namely, $\rho^{(\rm em)}\propto b^{-4}$, as the Universe expands and becomes almost constant for large values of $\alpha$ relevant to inflationary cosmologies namely, $\alpha\gtrsim 40$. For instance, taking $\alpha\sim40$, after 60 e-folds its value would drop only about ten times.
\\
\linebreak
We note first that the model exhibits qualitatively different behaviours at relatively earlier times depending on whether the $k$ is null or not, though the cases for $k\neq0$ eventually approaches the particular case $k=0$ at large $t$ values. First of all we would like to note that, in what follows, unless otherwise stated, we shall carry out our discussions by considering $\alpha>2$, and that we are particularly interested in large $\alpha$ values (say $\alpha \gtrsim 40$) in line with inflation paradigm\footnote{Although we are interested in the cases for large $\alpha$ values, the following two cases worth mentioning: (i) The particular case $k=0$ and $\alpha=\frac{2}{3}$ leads to an empty spacetime $\rho=0$, $p_x=p_y=p_z=0$, $Y=0$ with the following kinematical properties $R=0$, $\Delta=2$, $q=1$, $a\propto t^{-\frac{1}{3}}$ and $b\propto t^{\frac{2}{3}}$, $v\propto t$, which corresponds to a particular case of Kasner solution for LRS Bianchi I spacetime. According to this, $\alpha=\frac{2}{3}$ is a limiting case and the values of $\alpha$ less than $\frac{2}{3}$ would render the energy density of the effective fluid negative.  (ii) The particular case $k=0$ and $\alpha=\frac{1}{2}$ leads to $Y={\rm const.}$, $R=0$, EoS parameter of the standard electromagnetic field $[w_x,w_y,w_z]=[-1,1,1]$ as expected in the Einstein-Maxwell model. However, we note from \eqref{eqn:memr} that in this case $\rho^{(\rm mEMF)}$ becomes negative, implying that the mEMF could not be reduced to the standard electromagnetic field properly. It is easy to see that this is happening because of the excessively large value of the expansion anisotropy, which is $\Delta=8$, in this solution, due to the relation $3(1-\Delta /2)H^2=\kappa^2\rho$ for Bianchi type-I spacetimes in GR \cite{Akarsu:2010zm}. (Please see Section \ref{section:CR} for further discussion.) }. The condition $\alpha>2$ guarantees that all the axes of the Universe would eventually expand with increasing rate as can be seen from \eqref{kinx} and \eqref{kiny}.

In the general case $k\neq0$, the model exhibits a pancake like metric singularity at the beginning, namely, as $t\rightarrow 0$, the mean scale factor vanishes ($v\rightarrow 0$) but such that $a\rightarrow 0$ and $H_x\rightarrow \infty$ while $b\rightarrow v_1 k^{\frac{1}{3}}\alpha^{-\frac{1}{3}} t_{0}^{-\frac{1}{3}}$ and $H_{y,z}\rightarrow 0$. The energy density of the effective fluid vanishes at this limit, namely, $\rho\rightarrow 0$ as $v\rightarrow0$. The reason being that the effective fluid behaves like a phantom field, that is its average EoS parameter is below the phantom divide line (PDL) $-1$ and hence increases/decreases in energy density with increasing/decreasing volume as $\rho\propto v^{\frac{9\alpha-6}{3\alpha-1}}$ approximately, for the times earlier than $t_{\rm pdl}=\left(\frac{3\alpha \sqrt{3\alpha-1}-2(3\alpha-1)}{3\alpha-2}\right)^{\frac{1}{3\alpha}} k^{\frac{1}{3\alpha}} $ and hence its energy density vanishes as $v\rightarrow 0$. All these imply that there is no Big Bang singularity in the model provided that $k\neq0$. In the particular case $k=0$, on the other hand, there is a point like metric singularity accompanied by infinitely large energy density, hence a Big Bang like singularity, in the $t\rightarrow0$ limit, namely, $a\rightarrow 0$, $b\rightarrow 0$, $H_x\rightarrow \infty$, $H_{y,z}\rightarrow \infty$ and $\rho\rightarrow\infty$ as $t\rightarrow 0$.  Although the model exhibits completely different behaviour at very early times depending on whether the constant $k$ is null or not, all $k\neq 0$ cases eventually evolve into the case $k=0$.

\begin{figure}[t!]
     \begin{center}
        \subfigure[]{%
            \label{fig:HP}
            \includegraphics[width=0.31\textwidth]{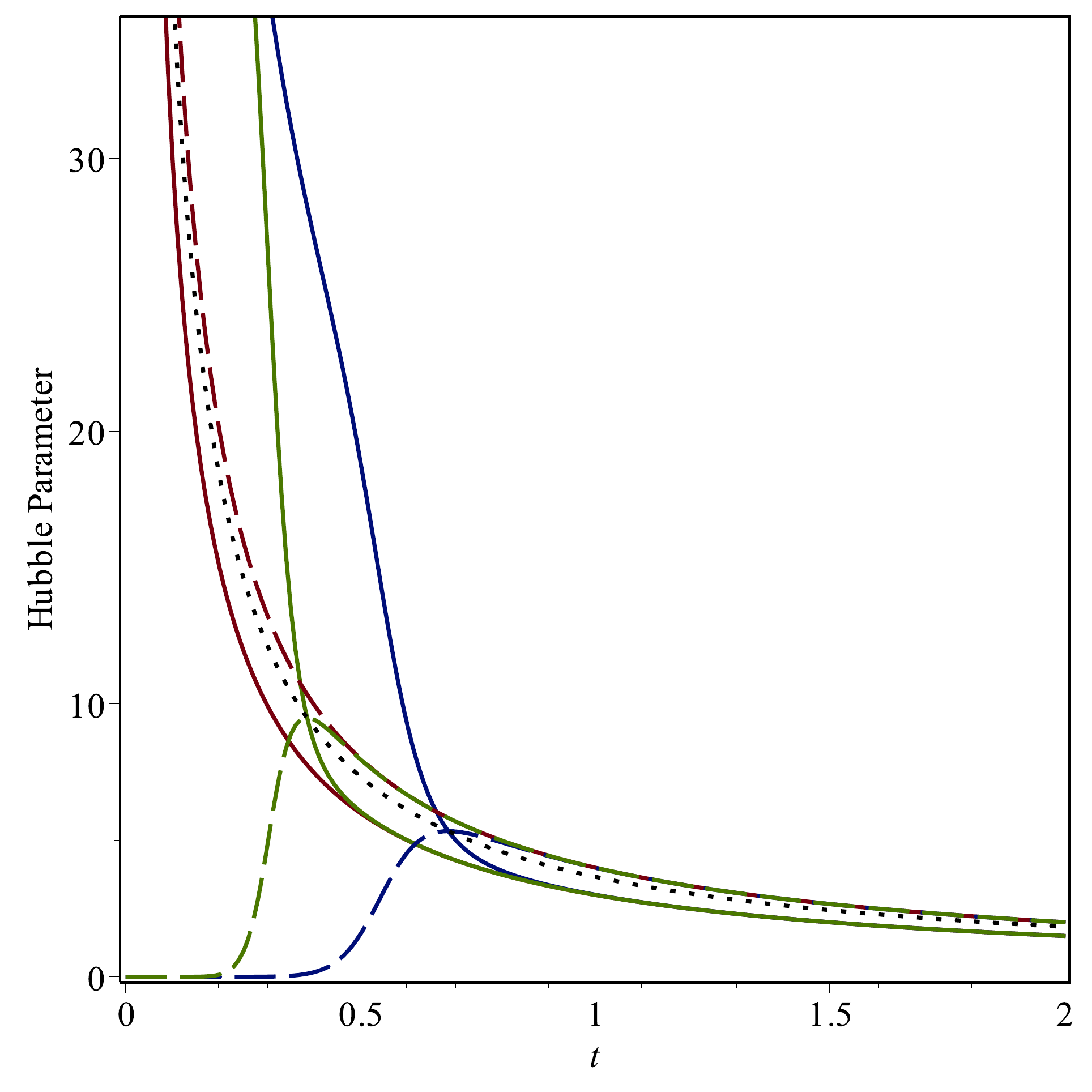}
        }%
        \subfigure[]{%
           \label{fig:deltaG}
           \includegraphics[width=0.31\textwidth]{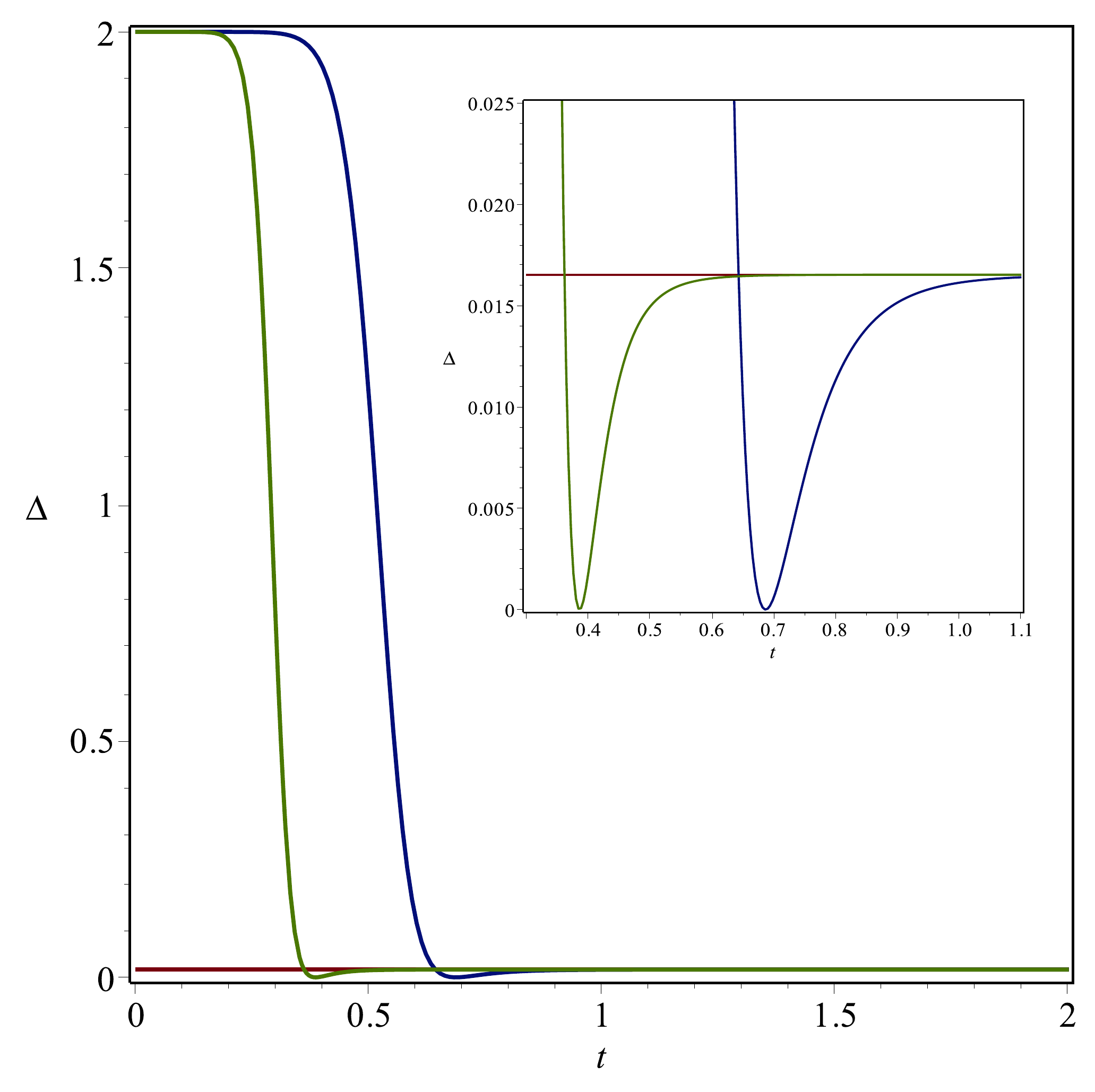}
        }
        \subfigure[]{%
            \label{fig:DPG}
            \includegraphics[width=0.31\textwidth]{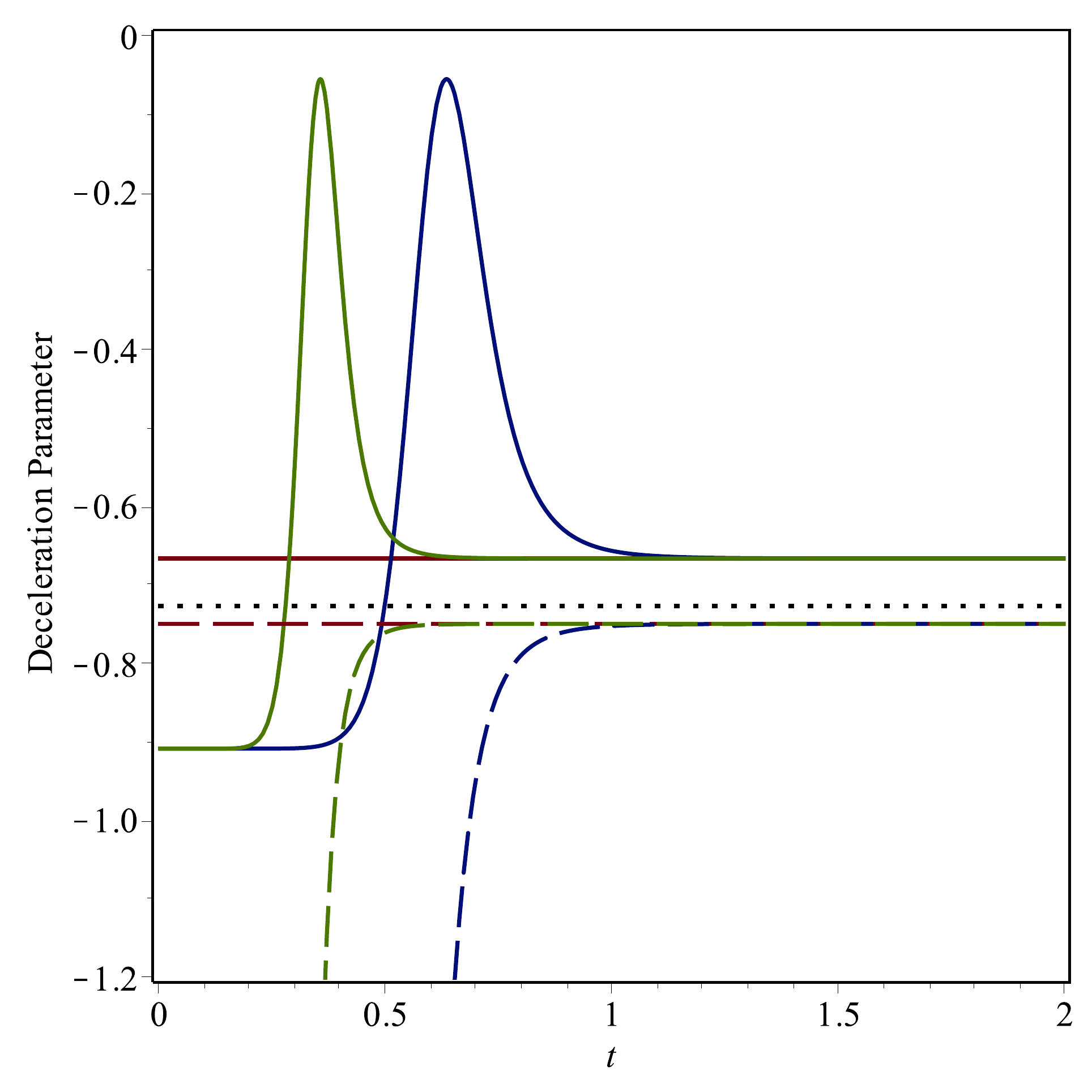}
        }%
        
    \end{center}
    \caption{%
       \textbf{(a)} Mean $H$ (dotted) and directional Hubble parameters $H_x$ (solid) and $H_{y,z}$ (dashed) versus cosmic time $t$. \textbf{(b)} Expansion anisotropy $\Delta$ versus cosmic time $t$. \textbf{(c)} Mean $q$ (dotted) and directional deceleration parameters $q_x$ (solid), $q_{y,z}$ (dashed) versus cosmic time $t$. In all figures the parameters are plotted by choosing $\alpha=4$ and $k=0$ (red), $k=10^{-6}$ (green) and $k=10^{-3}$ (blue).
     }%
	\label{fig:contours}
\end{figure}

We note that the expansion anisotropy $\Delta$ is a constant that ranges from 0 to 2 depending on $\alpha$ for $k=0$, while it is not only dynamical but also non-monotonic for all non zero $k$ values, such that, it starts at value 2, which implies the Universe is dominated by the expansion anisotropy as in the Kasner solution ($\rho=0$) that yields $\Delta=2$, then it vanishes at $t=[k(3\alpha-1)]^{\frac{1}{3\alpha}}$ and eventually approaches a constant value given in \eqref{anik0} depending on $\alpha$ asymptotically as $t$ increases. One may check from the directional Hubble and deceleration parameters that this non-monotonic behaviour of the expansion anisotropy is because the expansion rate of the $x$-axis, which was initially infinitely large, decreases ($\dot{H}_{x}<0$ and $q_x>-1$) and evolves below the expansion rate of the $y$- and $z$-axes, which was initially zero ($H_{y,z}(t=0)=0$) but increases due to a super-exponential expansion ($\dot{H}_{y,z}>0$ and $q_{y,z}<-1$) until $t=t_{\rm pdl}$. We demonstrate the characteristic behaviours of the mean and directional Hubble parameters in Fig. \ref{fig:HP}, of the expansion anisotropy in Fig. \ref{fig:deltaG} and of the mean and directional deceleration parameters in Fig \ref{fig:DPG} for $\alpha>1$ by using exaggerated values for the constants with the purpose of a better view. It is clearly seen from the figures that the cosmological parameters that behave nontrivially at relatively earlier times for the case $k\neq0$ eventually evolve into the simpler behaviours given in case $k=0$. Such a behaviour, in spite of the anisotropic expansion, may lead all axes to undergo the same amount of expansion, namely number of e-folds, during the inflationary era, as in the inflationary models constructed within the framework of RW spacetime relying on cosmic no hair theorem, but yet lead to a tiny anisotropy in the expansion of the Universe at the end of inflation. It is obvious that this is not possible in case $k=0$, which yields $\Delta={\rm const.}$, since, in this case, the expansion rate along the $x$-axis is always less than that of the $y$- and $z$-axes.

\begin{figure}
     \begin{center}
        \subfigure[]{%
            \label{fig:rho}
            \includegraphics[width=0.31\textwidth]{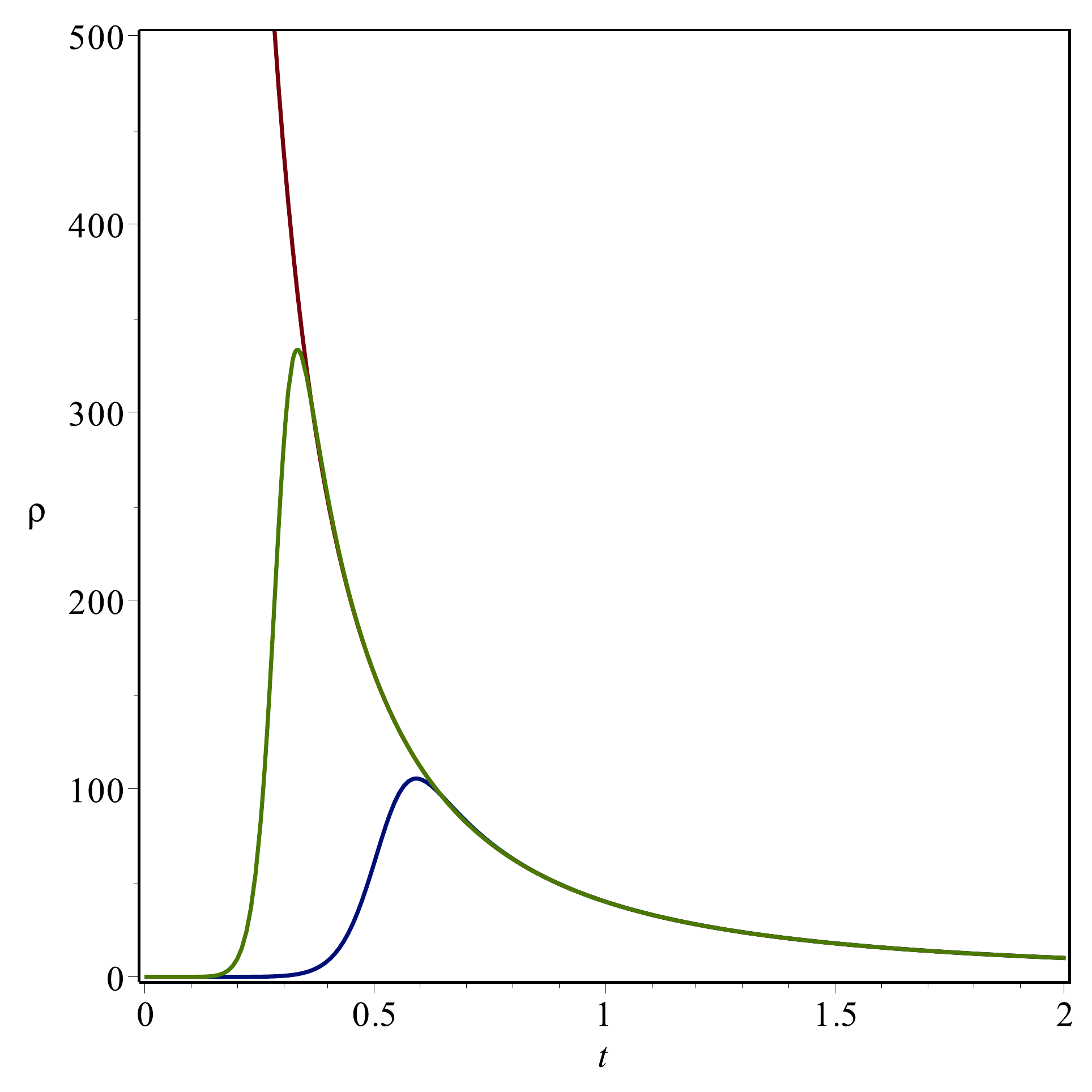}
        }%
        \subfigure[]{%
           \label{fig:eosp}
           \includegraphics[width=0.31\textwidth]{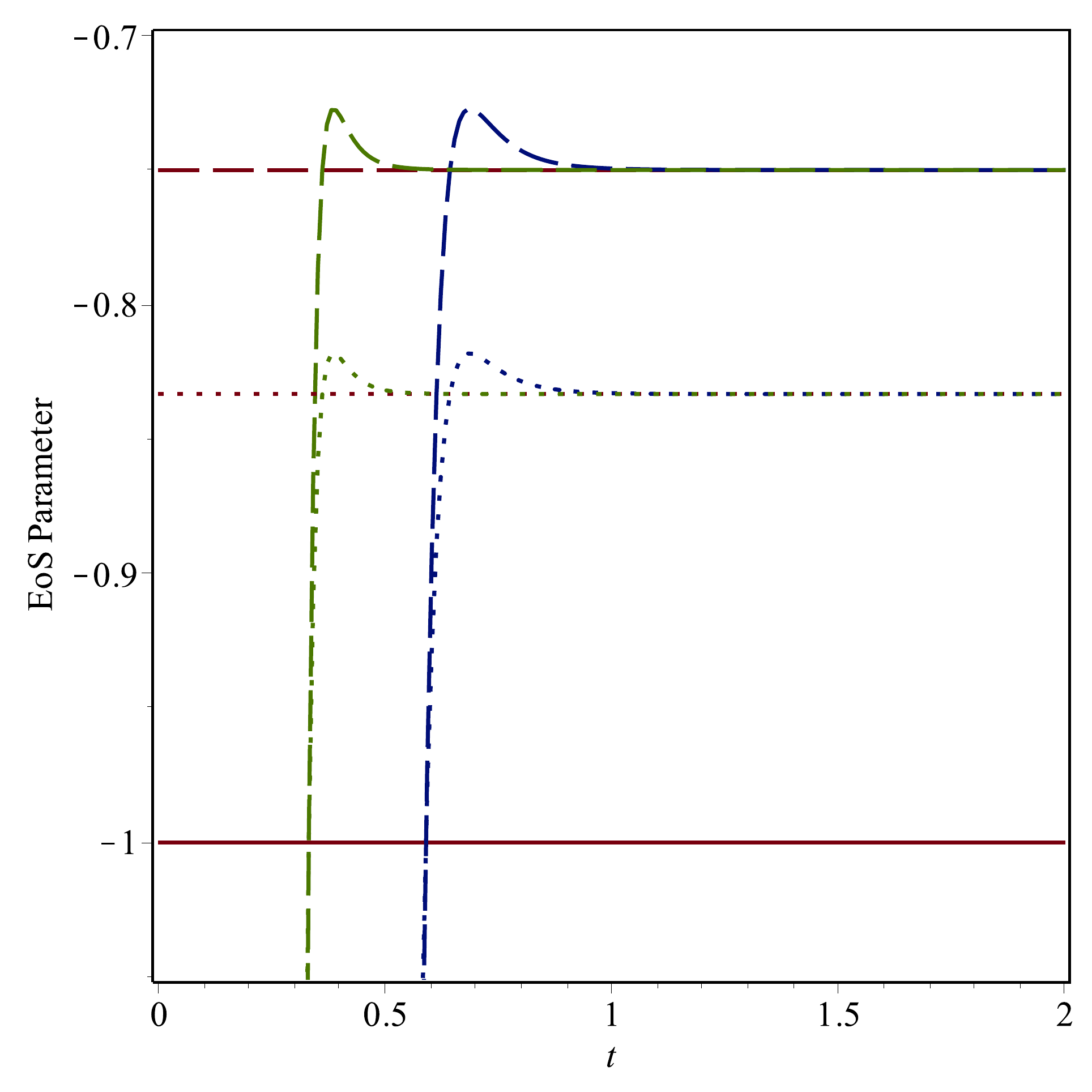}
        }%
        \subfigure[]{%
           \label{fig:Y}
           \includegraphics[width=0.31\textwidth]{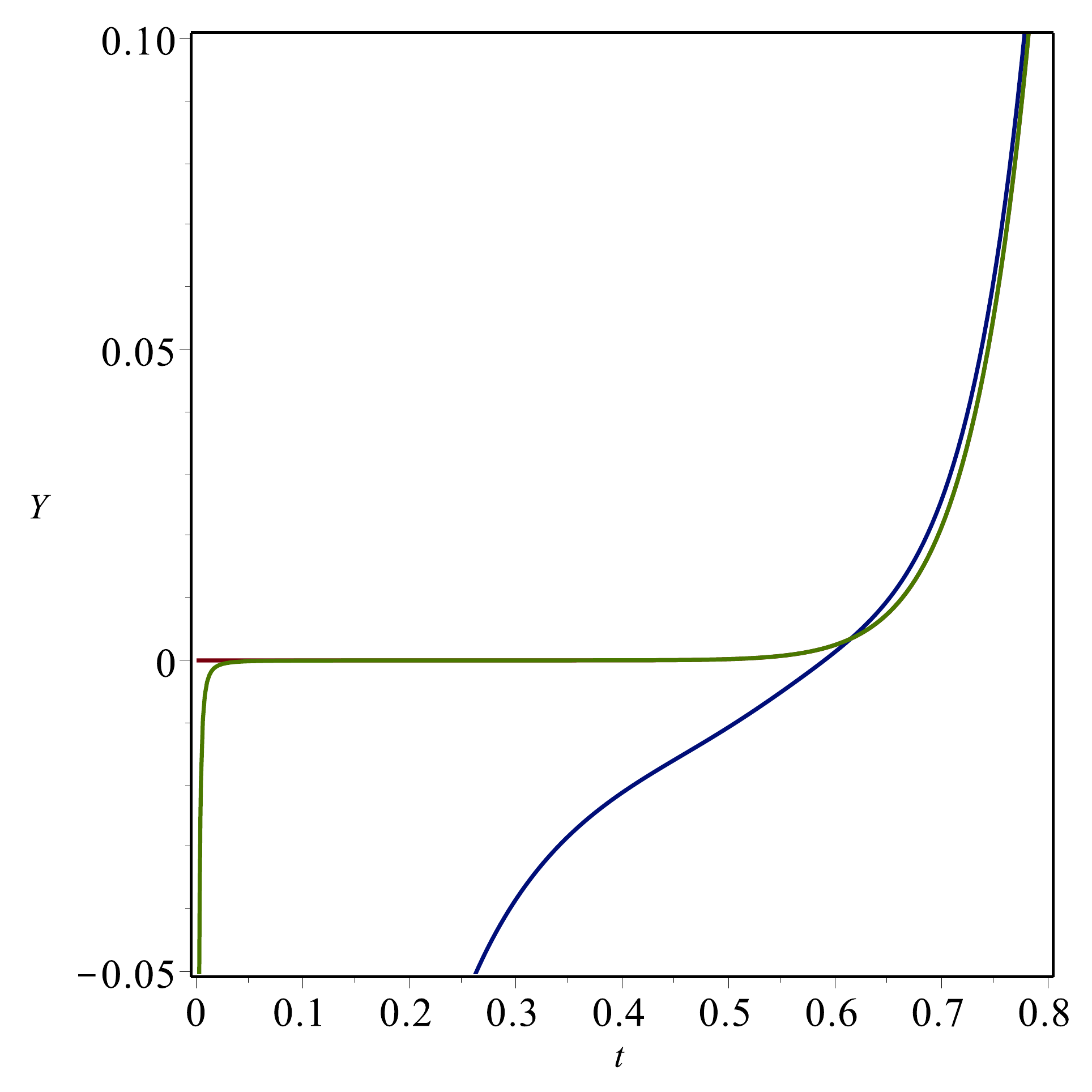}
        }%
                
    \end{center}
    \caption{%
       \textbf{(a)} Energy density $\rho$ versus cosmic time $t$. \textbf{(b)} The directional equation of state parameters versus cosmic time $t$. \textbf{(c)} $Y$ versus $t$. The parameters are plotted by choosing $k=0$ (red), $k=10^{-6}$ (green) and $k=10^{-3}$ (blue).
     }%
	\label{fig:contours}
\end{figure}

The energy density of the effective fluid $\rho$, for the case $k=0$, starts at infinitely large values at the beginning and decreases monotonically as $\rho\propto v^{-\frac{6}{3\alpha-1}}$. For the cases $k\neq0$, on the other hand, it behaves non-monotonically; such that, it is null at the beginning $t=0$, and increases as $t$ increases as $\rho\sim v^{\frac{9\alpha-6}{3\alpha-1}}$ in the period $t<t_{\rm pdl}$, during which the coupling function $Y<0$ leading the average EoS parameter to lie in the phantom region $w_{y,z}<w_x=-1$. It reaches its finite maximum value at $t=t_{\rm pdl}$, namely, when $Y=0$ leading $w_{y,z}=w_x=-1$. For the times $t>t_{\rm pdl}$, $Y>0$ leading $w_{y,z}>w_{x}=-1$ and it decreases monotonically as $\rho\sim v^{-\frac{6}{3\alpha-1}}$ with directional EoS parameters $w_x=-1$ and $w_{y,z}\sim-1+\frac{1}{\alpha}$, that is, approaches the solution for $\rho$ in case $k=0$. We note that any deviation from $w_{y,z}=-1$ implies anisotropic EoS parameter namely, $w_{y,z}$, which is dynamical for $k\neq0$, deviates from $-1$ while $w_{x}={\rm constant}=-1$. Therefore we have a dynamically anisotropic effective fluid, which leads nontrivial evolution of the expansion anisotropy as shown in Fig. \ref{fig:deltaG}. We demonstrate the characteristic behaviours of the energy density in Fig. \ref{fig:rho}, of the mean and directional EoS parameters in Fig. \ref{fig:eosp} and of the non-minimal coupling function $Y$ in \ref{fig:Y} by using exaggerated values for the constants with the purpose of a better view. 

The power-law inflation that we consider for investigating our gravity model, in fact, has a drawback namely the Universe accelerates forever and hence cannot accommodate deceleration in the form of the radiation dominated epoch which succeeds inflation. One may see \cite{Unnikrishnan:2013vga} and references therein about a possible way of dealing with this point and a further discussion as well on the current state of the power-law inflation in the light of Planck results. However, in this paper, we are interested in the expansion properties of the Universe during the inflationary era in which the Universe expands for about sixty e-folds, rather than a complete inflationary scenario with a graceful exit mechanism. Therefore, it would be useful to investigate the anisotropic e-fold, namely, the e-fold numbers along the $x$-axis and $y$- and $z$-axes separately with respect to the e-fold number of the mean scale factor. The e-fold number used to express the amount of inflation, by  considering the mean scale factor, can be given as
\begin{equation}
N=\ln{\frac{v_{\rm end}}{v}},
\end{equation}
where $v_{\rm end}$ is the mean scale factor at the end of inflation. Typically it is expected that the inflationary era lasts for  $\sim 60$ e-folds and that the CMB anisotropies correspond to perturbations whose wavelengths crossed the Hubble radius around $N_* \sim 60$ before the end of inflation. Here and after, $_*$ represents the value of a parameter  $N_*$ e-folds before the end of inflation. Given these, the inflation ends at the time $t=t_{\rm end}$  and number of e-folds is set to zero for that time as $N_{\rm end}=N(t_{\rm end})=0$, we find
\begin{equation}
\label{eqn:efdef}
t=t_{\rm end}\, {\rm e}^{-\frac{3N}{3\alpha-1}}
\end{equation}
which leads to
\begin{equation}
v=v_1 t_{\rm end}^{\alpha-\frac{1}{3}} \, {\rm e}^{-N}
\end{equation}
for the mean scale factor. Next using these we find the following relations between the e-folds of the mean scale factor and that of the directional scale factors in terms of the parameters $\alpha$ and $n=k\, t_{\rm end}^{-3\alpha}\, {\rm e}^{\frac{9\alpha N_*}{3\alpha-1}}$:
\begin{equation}
N_x=\ln{\frac{a_{\rm end}}{a}}= \frac{3\alpha-3}{3\alpha-1}\, N+\frac{2}{3}\ln\left[\frac{1+n {\rm e}^{-\frac{9\alpha (N_*-N)}{3\alpha-1}}}{1+n{\rm e}^{-\frac{9\alpha N_*}{3\alpha-1}}}\right]
\end{equation}
and
\begin{equation}
N_{y,z}=\ln{\frac{b_{\rm end}}{b}}= \frac{3\alpha}{3\alpha-1}\, N-\frac{1}{3}\ln\left[\frac{1+n {\rm e}^{-\frac{9\alpha (N_*-N)}{3\alpha-1}}}{1+n{\rm e}^{-\frac{9\alpha N_*}{3\alpha-1}}}\right].
\end{equation}
It follows that the difference between the e-folds of the directional scale factors is
\begin{equation}
N_{y,z}-N_x=\frac{3N}{3\alpha-1}-\ln\left[\frac{1+n {\rm e}^{-\frac{9\alpha (N_*-N)}{3\alpha-1}}}{1+n{\rm e}^{-\frac{9\alpha N_*}{3\alpha-1}}}\right].
\end{equation}
For the case $n=0$, these read
\begin{equation}
N_x=\frac{3\alpha-3}{3\alpha-1}\,N\quad\textnormal{and}\quad N_{y,z}= \frac{3\alpha}{3\alpha-1} \,N,
\end{equation}
and
\begin{equation}
N_{y,z}-N_x=\frac{3N}{3\alpha-1}.
\end{equation}
We can find the number of e-folds for the scale factors during the inflationary era by choosing $N=N_*$ in the above equations. Doing so, because $\alpha$ should typically be much larger than $1/3$, we see that the number of e-folds of the $y$- and $z$-axes is always greater than that of the $x$-axis in the case $n=0$, while it can be reduced and even set to zero by choosing the appropriate value for $n$. We find accordingly that, the e-folds for all the axes are the same,
\begin{equation}
N_{*x}=N_{*y,z} \quad\textnormal{if}\quad n=\frac{{\rm e}^{\frac{9\alpha N_*}{3\alpha-1}}-{\rm e}^{3N_*}}{{\rm e}^{3N_*}-1}.
\end{equation}

\begin{figure}[t]
     \begin{center}
        \subfigure[]{%
            \label{fig:efold}
            \includegraphics[width=0.45\textwidth]{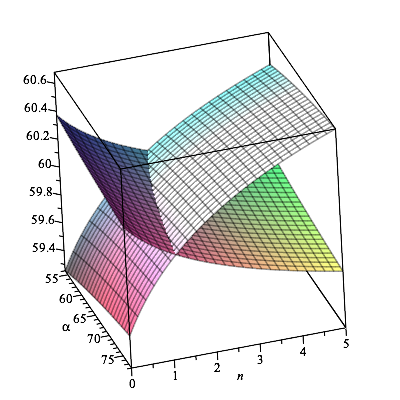}
        }%
        \subfigure[]{%
           \label{fig:defold}
           \includegraphics[width=0.45\textwidth]{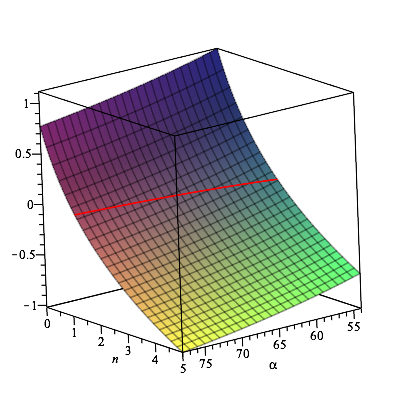}
        }%
        
    \end{center}
    \caption{%
       \textbf{(a)}$N_{*x}$ and $N_{*y,z}$, e-folds of the $x$-axes and $y$- and $z$-axes, versus $\alpha=[54,78]$ and $n=[0,5]$ by choosing the e-folds of mean scale factor $N^{*}=60$. \textbf{(b)} $N_{*y,z}-N_{*x}$, the difference between e-folds of the $x$-axes and $y$- and $z$-axes, versus $\alpha=[54,78]$ and $n=[0,5]$ by choosing the e-folds of mean scale factor $N^{*}=60$. The red curve represents $N_{*y,z}-N_{*x}=0$.
     }%
	\label{fig:contours}
\end{figure}

\begin{figure}[t]\centering
\includegraphics[width=0.45\textwidth]{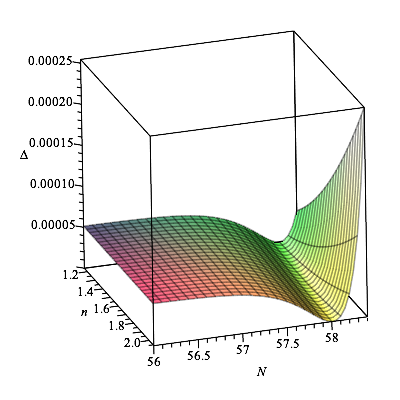}
\caption{$\Delta$ versus e-folds $N$ considering $n$ ranges from $2.06$ to $1.16$ relying on the recent observational value of spectral index from Planck 2015 results. $N$ decreases as the cosmic time $t$ increases.}
\label{fig:Deltann}
\end{figure}

In power-law inflation models based on RW spacetime, when slow roll is not assumed, scalar spectral index $n_s$, which measures the slight deviation from scale invariance, is given by $n_s=1-\frac{6}{3\alpha-4}$, independent of whether we consider canonical or non-canonical scalar field \cite{Unnikrishnan:2013vga}. Accordingly, the Planck full mission temperature data and a first release of polarisation data on large angular scales measure the spectral index of curvature perturbations to be $n_s = 0.968 \pm 0.006$ at $95\%$ CL \cite{Ade:2015lrj} that restricts $\alpha$ to the range
\begin{equation}
\label{eqn:alpharange}
54\lesssim \alpha \lesssim 78.
\end{equation}
Considering this range for the case $n=0$ in our model, we find the difference between the number of the e-folds of the directional scale factors as follows:
\begin{equation}
n=0 \quad \Rightarrow \quad0.77 \lesssim N_{*y,z}-N_{*x}\lesssim 1.12\quad\textnormal{for}\quad N_* \sim 60.
\end{equation}
Hence in the case $n=0$, in which not only the mean scale factor but also the directional scale factors exhibit power-law expansion, the $y$- and $z$-axes are approximately one e-fold ahead from the $x$-axis at the end of inflation. On the other hand, non-zero values of $n$ will certainly change this range. We depict the $N_{*x}$ and $N_{*y,z}$ surfaces in Fig \ref{fig:defold} and the $N_{*x}-N_{*y,z}$ surface in Fig \ref{fig:defold} for $N_*=60$ by considering the range for $\alpha$ given in \eqref{eqn:alpharange}. We note in Fig \ref{fig:efold} that there is an intersection of two surfaces, or the curve $N_{*x}-N_{*y,z}=0$ in Fig \ref{fig:defold}, implying all the axes undergo same amount of e-folds during the inflation, namely, $N_*=N_{*x}=N_{*y,z}=60$, which restricts the value of $n$ in the range
\begin{equation}
\label{eqn:rangen}
2.06>n>1.16
\end{equation}
for the range of $\alpha$ given in \eqref{eqn:alpharange}.
This of course comes out as a result of the non-monotonic behaviour of the expansion anisotropy: the expansion rate of the $y$- and $z$-axes, which was less than that of the $x$-axis at the beginning of the inflation, exceeds that of the $x$-axis at some point during the inflationary era and then $y$- and $z$-axes that were lagging behind in the earlier times of the inflation catch up the $x$-axis at the end of inflation. Thus we end up with an isotropic inflation, i.e., all the axes undergo the same number of e-folds during the inflation as in the inflationary models based on RW spacetime. However, this implies also that the expansion rate of the $y$- and $z$-axes should be larger than that of the $x$-axis at the end of inflation, and hence in contrast to the inflationary models based on RW spacetime there should be a residual expansion anisotropy at the end of inflation. We can write the expansion anisotropy in terms of e-folding of the mean scale factor as follows:
\begin{equation}
\label{eqn:Defold}
\Delta=\frac{2}{(3\alpha-1)^2}\, \left(1-\frac{3\alpha n}{n+{\rm e}^{\frac{9\alpha(N_*-N)}{3\alpha-1}} }     \right)^2.
\end{equation}
From this, we find that
\begin{equation}
\label{eqn:N058}
\Delta=0 \quad\textnormal{at}\quad N=\frac{3\alpha-1}{9\alpha} \ln \left( \frac{{\rm e}^{\frac{9\alpha N_*}{3\alpha-1}}}{n(3\alpha-1)}\right).
\end{equation}
Using the ranges for $\alpha$ and $n$ given in \eqref{eqn:alpharange} and \eqref{eqn:rangen}, respectively, we find that
\begin{equation}
\Delta=0\quad\textnormal{when}\quad N\sim 58
\end{equation}
that is, the expansion anisotropy vanishes approximately 58 e-folds before the end of inflationary era ($N=0$) during which all the axes underwent 60 e-folding, i.e., $N_*=N_{*x}=N_{*y,z}=60$. Its value at the end of inflation can be obtained by setting $N=0$ in \eqref{eqn:Defold} and it is easy to see that 
\begin{equation}
\Delta(N=0)=\Delta_{\rm end}\simeq\frac{2}{(3\alpha-1)^2}
\end{equation}
for the above considered values for the parameters $\alpha$, $N_*$ and $n$. Accordingly we find that our model predicts a residual anisotropy in the expansion rate of the Universe that ranges as
\begin{equation}
\label{eqn:Deltaendr}
7.73\times 10^{-5}>\Delta_{\rm end}>3.66 \times10^{-5}.
\end{equation}
We calculate from \eqref{eqn:Defold} that $\Delta$ reaches values approximately equal to $\Delta_{\rm end}\sim 5\times10^{-5}$ at $N\sim57$ after passing its minimum value equal to zero at $N\sim58$. Hence the range we give for $\Delta_{\rm end}$ in \eqref{eqn:Deltaendr} is in fact also the range of the expansion anisotropy during a large part of the inflationary era, namely, for about the last 57 e-folds of 60 e-folds. We depict in Fig \ref{fig:Deltann} the evolution of the expansion anisotropy $\Delta$ in terms number of e-folds $N$ considering the range given for $n$ in \eqref{eqn:rangen}.
This also shows that the dynamics of the Universe during the last $\sim57$ e-folds can be very well described by the particular case $k=0$ of our solution. Accordingly,  we find that the Universe undergoing a near de Sitter inflation exhibits a slightly anisotropic acceleration during the last $\sim57$ e-folds, such that the directional deceleration parameters differ at the fourth decimal place;
\begin{align}
\alpha=54\,\, (q=-0.9813)\quad \Rightarrow \quad  q_x=-0.9811 \quad\textnormal{and}\quad q_{y,z}=-0.9815\\
\alpha=78\,\, ( q=-0.9872) \quad \Rightarrow \quad  q_x=-0.9871 \quad\textnormal{and}\quad q_{y,z}=-0.9872.
\end{align}
As expected, during the last $\sim57$ e-folds where the Universe not only exhibits a near de Sitter inflation but also possesses a tiny expansion anisotropy, viz., $\Delta\lesssim10^{-4}$, the effective fluid behaves similar to the conventional vacuum energy with a constant energy density, namely, decreases very slowly as the Universe expands;
\begin{align}
\nonumber
\alpha=54\quad \Rightarrow \quad   \rho\propto v^{-0.037} \quad\textnormal{and}\quad \alpha=78\quad \Rightarrow \quad   \rho\propto v^{-0.026}.
\end{align}
This slight deviation of the energy density of the effective fluid from being constant is because the directional EoS parameter along the $y$- and $z$-axes takes slightly higher values than -1 while the one along the $x$-axis is constant equal to -1, namely,
\begin{align}
\nonumber
\alpha=54\quad \Rightarrow \quad   w_x=-1 \quad\textnormal{and}\quad w_{y,z}=-0.9815\\
\alpha=78\quad \Rightarrow \quad   w_x=-1 \quad\textnormal{and}\quad w_{y,z}=-0.9872.
\end{align}
This effective fluid, yielding a slightly anisotropic EoS parameter persistently during the last $\sim57$ e-folds of the inflation, avoids the expansion rate of the Universe to isotropize as the Universe expands and leads it to approach to a non-zero constant. On the other hand, considering the whole inflationary era, say all 60 e-folds, the expansion anisotropy is rapidly reduced to zero from its maximum value 2  in the first $\sim 3$ e-folds, and then a tiny expansion anisotropy is generated in the remaining $\sim57$ e-folds, and this non-monotic behavior can cause Universe to undergo the same amount of e-folds in all directions as in the isotropic inflationary models based on isotropic RW metric but leaves a tiny anisotropy in the expansion rate of the Universe, which would decay monotonically during the post-inflationary era, namely, in the radiation dominated era, so as to preserve the success of the standard Big Bang nucleosynthesis.

\begin{figure}
     \begin{center}
        \subfigure[]{%
            \label{fig:rhoN}
            \includegraphics[width=0.31\textwidth]{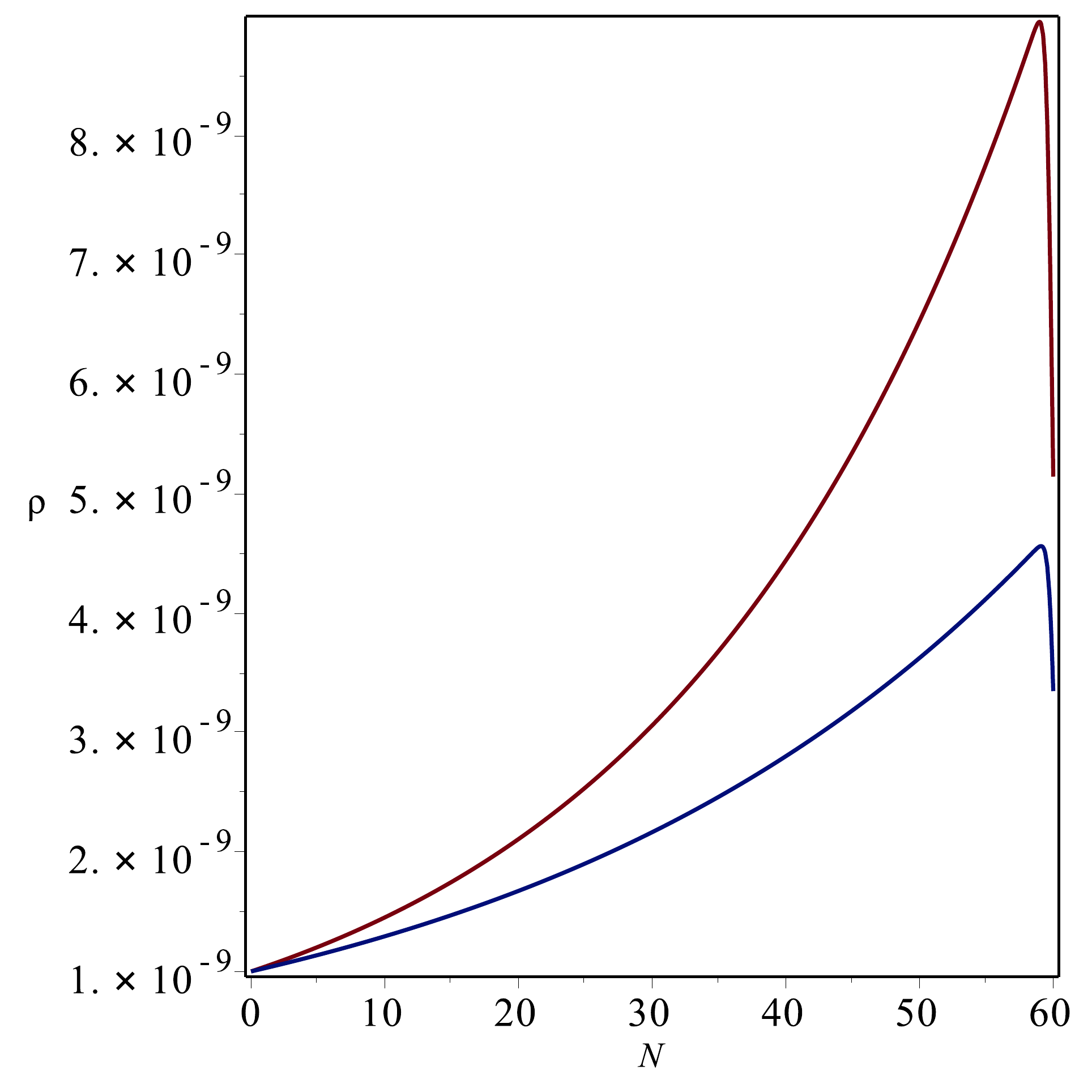}
        }%
        \subfigure[]{%
           \label{fig:rhoMN}
           \includegraphics[width=0.31\textwidth]{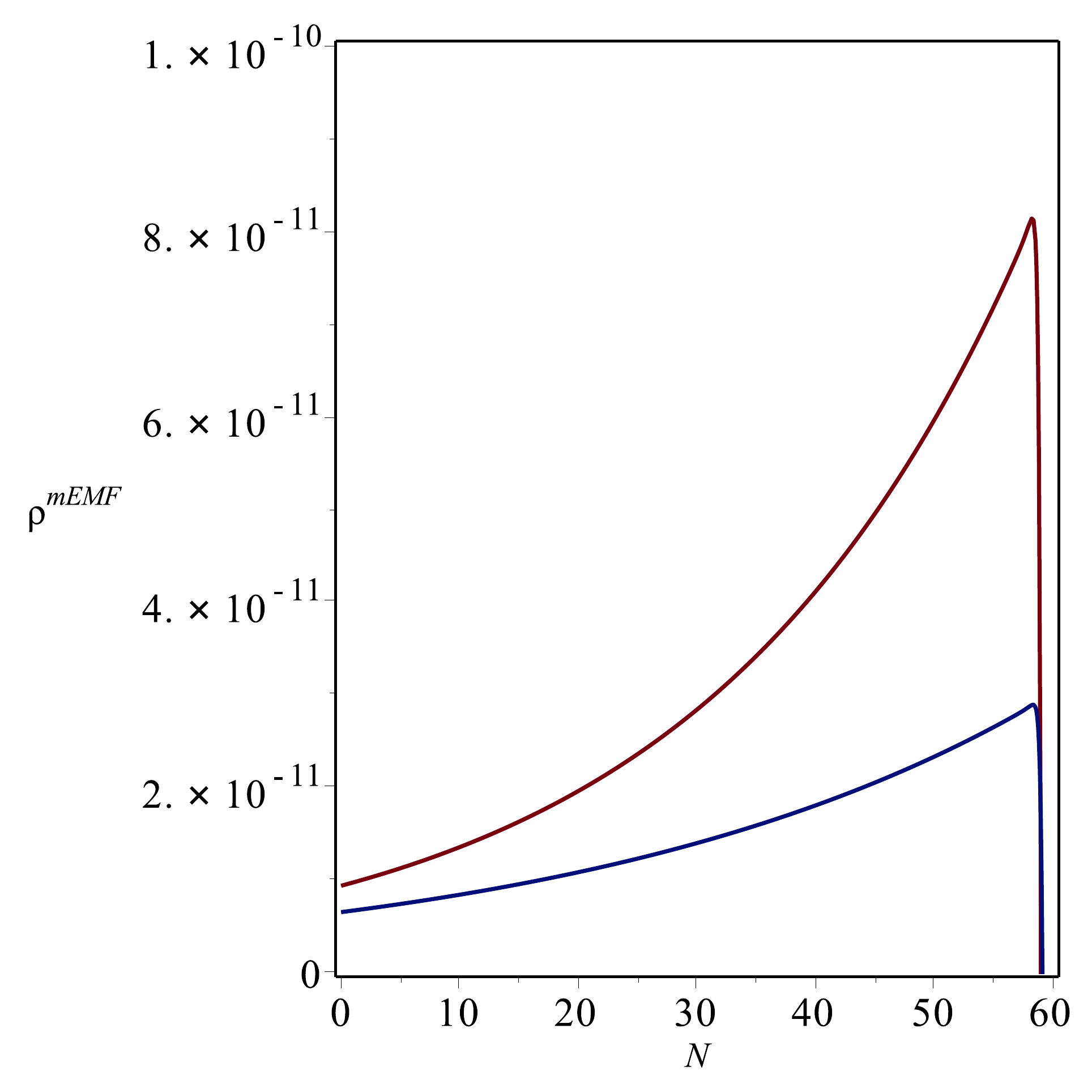}
        }%
        \subfigure[]{%
           \label{fig:ratioN}
           \includegraphics[width=0.31\textwidth]{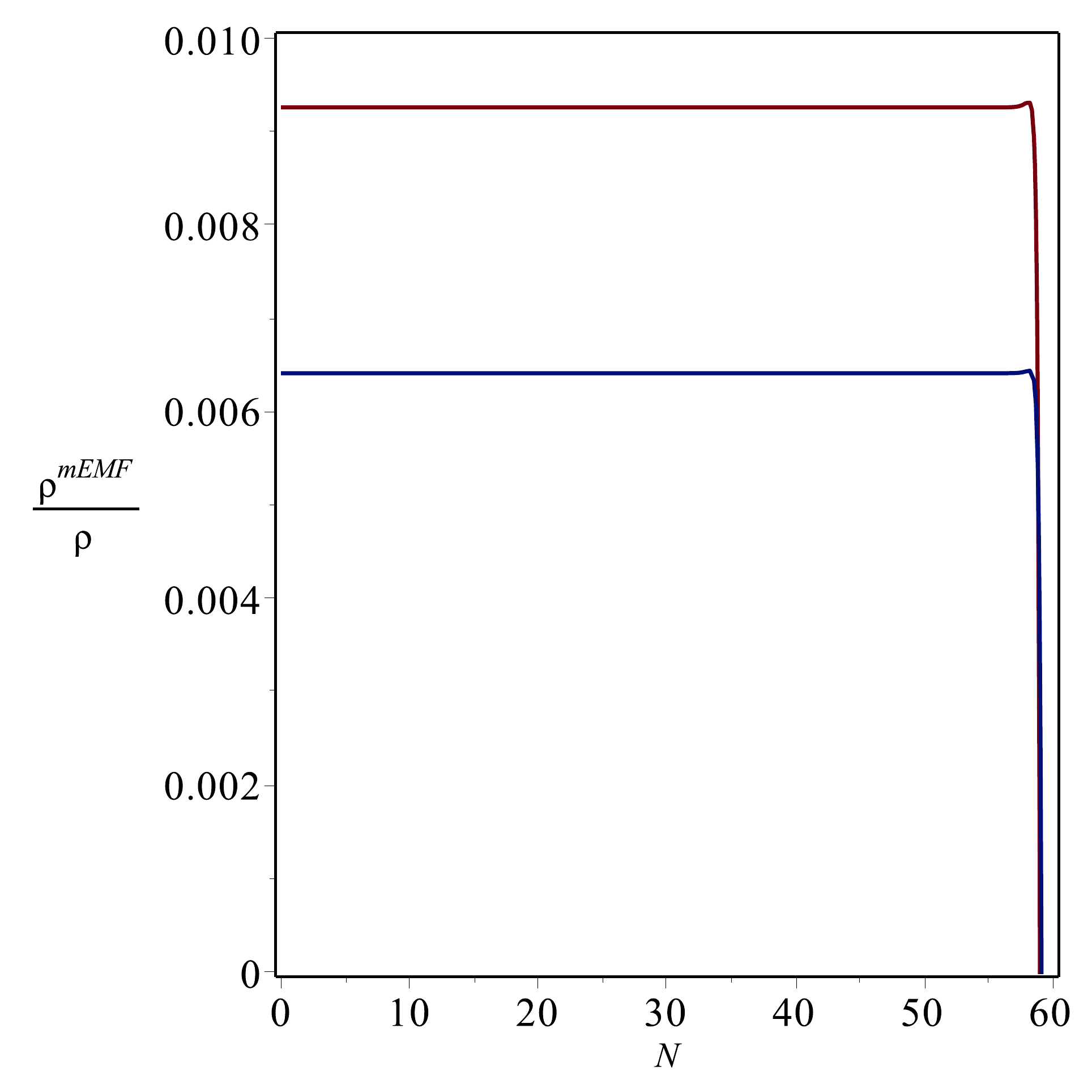}
        }%
                
    \end{center}
    \caption{%
      \textbf{(a)} Energy density of the effective fluid $\rho=\rho^{(\rm mEMF)}+\rho^{(\rm vac)}$ (total energy density) versus the mean scale factor e-fold $N$. \textbf{(b)} Energy density of the modified electromagnetic field $\rho^{(\rm mEMF)}$ versus the mean scale factor e-fold $N$. \textbf{(c)} Ratio of the energy density of the modified electromagnetic field to the total energy density $\rho^{(\rm mEMF)}/\rho$ versus the mean scale factor e-fold $N$. We set $\rho=10^{-9}\, {m}_{\rm Pl}^4$ at the end of inflation ($N=0$), and $N_*=60.$
     }%
	\label{fig:MEM}
\end{figure}

As the final task, we discuss on how the mEMF is generated from vacuum energy during the first few e-folds of the inflation and remains persistent afterwards against to the vacuum energy till end of inflation. To do so, we first write the energy densities of the mEMF and the effective fluid in terms of the e-folds of mean scale factor $N$ as follows;
\begin{equation}
\label{eqn:rhoMN}
\rho^{(\rm mEMF)} = \left(1+\frac{3\alpha}{3\alpha-2}\frac{2 n\, {\rm e}^{-\frac{9\alpha (N-N_*)}{3\alpha-1}}-(3\alpha-2) n^2}{ ({\rm e}^{-\frac{9\alpha (N-N_*)}{3\alpha-1}}+n)^2}\right) \, \frac{3\alpha -2  }{2\kappa^2\,t_{\rm end}^{2} }\, \, {\rm e}^{\frac{6N}{3\alpha-1}},
\end{equation}
\begin{equation}
\label{eqn:rhoN}
\rho= \left(1+\frac{3\alpha n}{3\alpha-2} {\rm e}^{\frac{9\alpha (N-N_*)}{3\alpha-1}} \right) \frac{{\rm e}^{\frac{9\alpha (N-N_*)}{3\alpha-1}}}{{\rm e}^{\frac{9\alpha (N-N_*)}{3\alpha-1}}+n} \, \frac{\alpha(3\alpha -2) }{\kappa^2\,t_{\rm end}^{2} }\, \, {\rm e}^{\frac{6N}{3\alpha-1}}.
\end{equation}
We note that for $N=0$ they approximate to the particular case $n=0$ ($k=0$), which read
\begin{equation}
\label{eqn:memvacN}
\rho^{(\rm mEMF)}=\frac{3\alpha -2  }{2\kappa^2\,t_{\rm end}^{2} }\, \, {\rm e}^{\frac{6N}{3\alpha-1}}\quad\textnormal{and}\quad \rho=\frac{\alpha(3\alpha -2) }{\kappa^2\,t_{\rm end}^{2} }\, \,\left(2\alpha-1\right) {\rm e}^{\frac{6N}{3\alpha-1}}.
\end{equation}
According to this, we find that
\begin{equation}
\frac{\rho^{(\rm mEMF)}}{\rho}\sim\frac{1}{2\alpha} \quad\textnormal{when}\quad N=0,
\end{equation}
i.e., at the end of inflation.
We see that for the case $n=0$ the energy densities decrease exponentially as the Universe expands (as $N$ decreases). We can immediately see, by choosing $n=0$, $\alpha=60$ and $N_*\sim60$, that $\rho^{(\rm mEMF)}_{\rm end}/\rho^{(\rm mEMF)}_{\rm N_*}\sim 0.1$, whereas the energy density of the standard electromagnetic field would decrease ${\rm e}^{240}$ times, and that $\frac{\rho^{(\rm mEMF)}}{\rho}\sim0.1$, the energy density of the electromagnetic field is persistent as the one percent of the total energy density of the Universe throughout the inflationary era. On the other hand, we see from \eqref{eqn:rhoMN} and \eqref{eqn:rhoN} that, for $n\neq0$, the energy densities of the mEMF and effective fluid (total energy density) evolve non-trivially. If we consider the values we obtained above, $2.06>n>1.16$ and $54\lesssim \alpha \lesssim 78$ for $N_*\sim60$, that lead all the axis of the Universe to experience the same number of e-folds by the end of inflation, we find that $\rho^{(\rm mEMF)}$, $\rho$ as well as $\frac{\rho^{(\rm mEMF)}}{\rho}$ evolves non-monotonically during the inflation. We depict in Figure \ref{fig:MEM} their evolution by setting the total energy density of the Universe at the end of inflation to a typical energy density value relevant to inflation as $10^{-9}\,m_{\rm Pl}^4$, where $m_{\rm Pl}\simeq1.22\times10^{19}\,{\rm GeV}$ is the Planck mass. We see from Figures \ref{fig:rhoMN} and \ref{fig:ratioN} that the mEMF is being rapidly generated from vacuum energy during the first few e-foldings ($N\sim60$), namely, both $\rho^{(\rm mEMF)}$ and $\frac{\rho^{(\rm mEMF)}}{\rho}$ increase and reache their maximum values. Afterwards, the energy density of the mEMF starts to decrease by keeping the ratio of its energy density to the energy density of the vacuum energy almost constant. By the end of inflation ($N=0$), the energy density of the mEMF becomes $\rho^{(\rm mEMF)}\sim10^{-11}\,m_{\rm Pl}^4$ and holds about the one percent of the total energy budget of the Universe, $\frac{\rho^{(\rm mEMF)}}{\rho}\sim 0.01$ at $N=0$. On the other hand, in the presence of standard electromagnetic field and vacuum energy, accelerated expansion can start when the energy densities of the electromagnetic field and vacuum energy are about the same value, and hence from the onset of the inflation to the end of inflation, after $60$ e-folds, the energy density of the standard electromagnetic field would hold only about the ${\rm e}^{-241}$ of the energy budget of the Universe.

\section{Closing remarks}
\label{section:CR}

We consider the non-minimal coupling of electromagnetic field to gravity in $Y(R) F^2$-form, where $F$ is the electromagnetic field and $Y(R)$ is a function of the scalar curvature $R$. We investigate a particular case of the model, for which the higher order derivatives that make the model too complicated for a general analytical investigation are eliminated but we keep $R$ to be dynamical via the constraint $Y_RF_{mn}F^{mn} =-\frac{2}{\kappa^2}$ \eqref{conditionX}. We show that this particular model can conveniently be elaborated either in the context of the trace-free Einstein/unimodular gravity or the GR. In fact, our model is equivalent to GR in the presence of an effective cosmological fluid consisting of a mEMF, viz. conventional electromagnetic field scaled by a factor $\frac{Y(R)}{2}$, and vacuum energy with a particular energy density $\rho^{(\rm vac)}=\frac{R}{4\kappa^2}$, non-minimally coupled through a dynamical factor $Y(R)$. In the context of cosmology, the mEMF here could be persistent by gaining energy-momentum from vacuum energy so that its anisotropic pressure can still play the main role in controlling the evolution of the expansion anisotropy and could even maintain an anisotropic expansion in spite of the inflation driven by vacuum energy.

It is well known that non-minimally coupled gravitational and electromagnetic fields yield complicated field equations which in general may involve higher derivatives that are known to produce generically ghost degrees of freedom, i.e., to cause the theory to be unstable as a particular consequence of Ostrogradski theorem (see e.g. \cite{EspositoFarese:2009aj} and references therein). We note that our Lagrangian \eqref{lag1} does not seem to be in the class of the so-called generalised Proca theories \cite{Heisenberg:2014rta,Jimenez:2016isa} that concern the generic vector-tensor actions that yield second order equations of motion, i.e., those theories free of Ostrogradski instabilities. In our study here, we eliminate the higher order derivatives by a constant (see \eqref{conditionX}) and reach the field equations \eqref{einstein2} from which we develop the rest of the work, as well as the cosmological application of the model that we carried out under the power-law volumetric expansion assumption \eqref{plvol}. For instance, we show in \eqref{YBRcheck} that by setting the relevant integration constant to zero in our exact solution, we recover $\frac{R^\eta}{M^{2\eta}}F^2$-form gravity with a particular coefficient, which is one of the most widely considered generalisation of Einstein-Maxwell theory (see e.g. \cite{Horndeski:1976gi,Turner:1987bw,MuellerHoissen:1988bp,Balakin:2010ar,Dereli:2011hu,Baykal:2015paa,Mazzitelli:1995mp,Bamba:2008ja,Campanelli:2008qp,Lambiase:2008zz,Kunze:2009bs,Dereli:2011gh}) especially for the generation of seed magnetic fields during the inflation \cite{Turner:1987bw,Mazzitelli:1995mp,Campanelli:2008qp,Lambiase:2008zz}. 

We then investigate an anisotropic cosmological model based on this gravity model and aim mostly at showing an interesting class of anisotropic inflation relying on $Y(R) F^2$-form gravity. We show that the non-monotically evolving expansion anisotropy that arises upon deviation from de Sitter expansion can lead the Universe to undergo the same amount of e-foldings in all directions by the end of inflation. We discuss that such a behavior owes its existence to the effective cosmological fluid of the model with a dynamically anisotropic EoS. Namely, its EoS parameter along the $x$-axis is exactly equal to $-1$, but is dynamical along the $y$- and $z$-axes such that it behaves as that of scalar fields provided that $Y>0$. Moreover, it can even cross below the phantom divide line provided that $Y<0$. We further study the near-de Sitter power-law inflation by assuming that the volume scale factor, rather than the directional scale factors, of the simplest anisotropic background metric is exhibiting a power-law expansion. We should note here that a thorough discussion on the inflationary observables of the model in the light of observational data would require the consideration of cases beyond the simplest power-law volumetric expansion, i.e., more featured forms of volumetric expansion law or suitably chosen forms of $Y(R)$ function instead. On the other hand, the exact solution we present under the power-law volumetric expansion assumption, as it is the simplest example for deviation from de Sitter expansion, provides us opportunity to demonstrate some of the features of such a scenario. It turns out that a {\it power-law volumetric expansion} does not necessarily imply directional scale factors exhibiting power-law expansion, which give rise to a constant expansion anisotropy, but leads to non-trivially evolving directional scale factors that give rise to a non-monotonically evolving expansion anisotropy that eventually converges to a non-zero constant. Relying on this fact, we direct our attention to the anisotropic e-fold during the inflation by considering the latest observational value of the spectral index of curvature from the Planck data and demanding the Universe to undergo the same amount of e-folds in all directions as in the isotropic inflationary models based on RW spacetimes. We show that in the inflationary era, which would last for about $60$ e-folds, the expansion anisotropy is rapidly reduced to zero from about its maximum possible value 2 in the first  $\sim3$ e-folds, and then only a small expansion anisotropy is generated in the remaining  $\sim57$ e-folds and an amount of $\Delta_{\rm end}\sim 10^{-4}$ anisotropy in the expansion rate of the Universe would remain at the end of inflation. This amount of anisotropy in the expansion would decrease to even smaller values monotonically during the post-inflationary era namely, in the radiation dominated era, so that the success of the standard Big Bang nucleosynthesis would be preserved. Such a scenario relies on the dynamically anisotropic character of the effective fluid that arises in our model.

It is showed that the mEMF is rapidly generated from vacuum energy during the first few e-folds of inflation, i.e., the energy of the mEMF not only increases in its value and reaches a maximum but also with respect to the total energy density of the Universe. Afterwards, the energy density of the mEMF starts to decrease by keeping the ratio of its energy density to the energy density of the effective fluid almost constant. By the end of inflation ($N=0$), the energy density of the mEMF becomes $\rho^{(\rm mEMF)}\sim10^{-11}\,m_{\rm Pl}^4$ and holds about the one percent of the total energy budget of the Universe, $\frac{\rho^{(\rm mEMF)}}{\rho}\sim 0.01$ at $N=0$. On the other hand, in the presence of standard electromagnetic field and vacuum energy, accelerated expansion can start when the energy densities of the electromagnetic field and vacuum energy are about the same value, and hence from the onset of the inflation to the end of inflation, after $60$ e-folds, the energy density of the standard electromagnetic field would hold only about the ${\rm e}^{-241}$ of the energy budget of the Universe. Assuming that a mechanism that switches the inflation off and leads our model to recover GR and standard electromagnetic field in the post-inflationary Universe could be found, this amount of electromagnetic field would be of astrophysical interest. Indeed, the ubiquitous presence of large-scale cosmological magnetic fields, not only in all astrophysical objects such as galaxies but also in voids, is still an unexplained feature of the Universe and the most promising answer so far is that such fields are relics from inflation, as first suggested in \cite{Turner:1987bw}. However, as it is well known, the conformal invariance of electromagnetism must be broken to render electromagnetic field persistent during inflation, which can be achieved by non-minimal coupling of electromagnetic field to gravity (see \cite{Turner:1987bw} for number of ways doing this) as we have done in this paper, or by non-minimal coupling of electromagnetic field to a scalar field as suggested first in \cite{Ratra:1991bn}. In \cite{Turner:1987bw}, the authors study $R F^2$-form non-minimal gravity along with a number of other ways of breaking the conformal invariance explicitly through gravitational couplings. Then it was argued first in \cite{Mazzitelli:1995mp} that its extension to $R^\eta F^2$-form non-minimal gravity, where $\eta$ is a positive integer, can generate seed electromagnetic field during inflation for adequate values of $\eta$ that would be of astrophysical interest (for more recent studies see, e.g., \cite{Campanelli:2008qp,Lambiase:2008zz} and references therein). We show that $\frac{R^\eta}{M^{2\eta}} F^2$-form non-minimal coupling is admitted in the particular form $\frac{\eta}{M^{2\eta}} R^{\eta-1}  F_{mn}F^{mn}=-\frac{2}{\kappa^2}$ by our model, which is however characterised by the constraint $Y_R F^2=-2/\kappa^2$ that eliminates higher order derivates in the field equations and render scalar curvature $R$ variable. However, our model demands negative $\eta$ values that are not necessarily integers, while it is assumed that $\eta$ is a positive integer in \cite{Turner:1987bw,Mazzitelli:1995mp,Campanelli:2008qp,Lambiase:2008zz}. The reason being that, in contrast to the works \cite{Turner:1987bw,Mazzitelli:1995mp,Campanelli:2008qp,Lambiase:2008zz} in this paper, we study near de Sitter inflation ($R>0$) in the presence of magnetic field ($F_{mn}F^{mn}>0$) only. Hence, it is of further interest for us to carry out a thorough investigation of whether inflationary cosmology base on the non-minimal model of gravity in $YF^2$-form characterised by the constraint $Y_R F^2=-2/\kappa^2$ could successfully account for large scale magnetic fields we observe today as done in \cite{Turner:1987bw,Mazzitelli:1995mp,Campanelli:2008qp,Lambiase:2008zz} in a separate study.

The dynamics for an inflationary cosmology in our model is drastically different than the dynamics of the Universe in the presence of standard electromagnetic field in GR (Einstein-Maxwell model), implying, for instance, a drastic deviation from the conformal invariance of the standard Maxwell theory, though nature shows no sign for this. Hence, to be able to consider the present non-minimal model of gravity in $Y F^2$-form characterised by the constraint $Y_R F^2=-2/\kappa^2$ as a viable model, we must show at least that the post-inflationary model that recovers or at least approximates to the general relativistic cosmology in the presence of standard electromagnetic field could be constructed in our model. Namely, standard Maxwell theory should be recovered or approximated at least after the electro-weak scale and expansion of the Universe should not deviate much from the predictions of GR, so that, for instance, the standard Big Bang Nucleosynthesis would not be spoiled. We discuss in Section \ref{model} that the choice $Y=1$ reduces our model to the Einstein-Maxwell model. Hence, one may then think that our model constrained by $Y_R F^2=-2/\kappa^2$ can approximate to the Einstein-Maxwell model, provided that $Y(R)$ function varies slowly enough around $Y=1$. However, as we have shown also in the exact solution obtained under the power-law volumetric expansion assumption that the Einstein-Maxwell model could be recovered but the energy density of the electromagnetic field appears with a wrong sign. We comment in Section \ref{sec:gen} also that this issue arises due to the excessively large value of the expansion anisotropy. Hence, a construction of our model from the start in a consistent way with the Robertson-Walker spacetime metric would set the expansion anisotropy to zero and then this issue would not arise. It is well known that a single electromagnetic field is incompatible with Robertson-Walker spacetime so that we consider the simplest anisotropic spacetime metric that can accommodate single electromagnetic field. On the other hand, isotropy of the electromagnetic field condensate can be achieved either in the case of a triplet of mutually orthogonal decoupled vector fields with equal lengths (dubbed as cosmic triad) or by considering a large number of randomly oriented decoupled vector fields, and in such a construction the electromagnetic field would behave like disordered radiation with an equation of state parameter equal to one third \cite{Bento:1992wy,ArmendarizPicon:2004pm,Golovnev:2008cf}. Hence, it would be interesting to further study cosmologies based on the non-minimal model of gravity in $Y F ^2$-form under the condition $Y_R F^2=-2/\kappa^2$ in such a setting and try to constrain the time rate of change of the $Y$, e.g., from Big Bang Nucleosynthesis. Indeed, we note from the field equations of the model \eqref{einstein2} that the function $Y$ would effect the expansion rate of the Universe just like a variable gravitational coupling (see, e.g., \cite{Copi:2003xd}) and hence it is conceivable that we can use the constraints on the expansion rate of the Universe during the time of BBN in this purpose.

\section*{Acknowledgements}
\"{O}.A. acknowledges the support by the Science Academy in the scheme of Distinguished Young Scientist Awards  (BAGEP). \"{O}.A. acknowledges further the financial support he received from, and hospitality of Koç University and the Abdus Salam International Centre for Theoretical Physics (ICTP), where parts of this work were carried out. \"{O}.S. acknowledges the support from Pamukkale University Scientific Research Fund with BAP project no: 2017HZDP009.

\end{document}